# Ultra-broadband reflectionless Brewster absorber protected by reciprocity

Jie Luo[1*#], Hongchen Chu[2#], Ruwen Peng[2], Mu Wang[2], Jensen Li[3,4*], Yun Lai[2*]

[1]School of Physical Science and Technology, Soochow University, Suzhou 215006, China

[2]National Laboratory of Solid State Microstructures, School of Physics, and Collaborative Innovation Center of Advanced Microstructures, Nanjing University, Nanjing 210093, China

[3]Department of Physics, The Hong Kong University of Science and Technology, Clear Water Bay, Hong Kong, China

[4]William Mong Institute of Nano Science and Technology, the Hong Kong University of Science and Technology, Clear Water Bay, Kowloon, Hong Kong, China

*Corresponding authors: Yun Lai (laiyun@nju.edu.cn); Jie Luo (luojie@suda.edu.cn); Jensen Li (jensenli@ust.hk)

#These authors contributed equally: Jie Luo, Hongchen Chu

**Abstract**

The Brewster's law predicts zero reflection of p-polarization on a dielectric surface at a particular angle. However, when loss is introduced into the permittivity of the dielectric, the Brewster condition breaks down and reflection unavoidably appears. In this work, we found an exception to this long-standing dilemma by creating a class of nonmagnetic anisotropic metamaterials, where an anomalous Brewster effects with independently tunable absorption and refraction emerges. This loss-independent Brewster effect is bestowed by the extra degrees of freedoms introduced by anisotropy and strictly protected by the reciprocity principle. The bandwidth can cover an extremely wide spectrum from dc to optical frequencies. Two examples of reflectionless Brewster absorbers with different Brewster angles are both demonstrated to achieve large absorbance in a wide spectrum via microwave experiments. Our work extends the scope of Brewster effect to the horizon of nonmagnetic absorptive materials, which promises an unprecedented wide bandwidth for reflectionless absorption with high efficiency.

**Introduction**

In the early 1810's, Sir David Brewster experimentally showed that when unpolarized light impinges on a dielectric interface, the reflected light can be linearly polarized if the reflected beam is perpendicular to the refracted one[1,2]. The origin of this Brewster effect lies in the elimination of reflection that occurs for the transverse-magnetic (TM)-polarized light at the Brewster's angle. However, when loss is introduced to the permittivity of dielectrics, a phase difference between the electric and magnetic fields appears, which introduces substantial reflection and breaks the Brewster condition[3,4]. Consequently, for the purpose of reflectionless absorption, complex engineering on the material permeability and dispersions[5,6] have been generally applied, which, however, introduces tremendous difficulty in the widening of the bandwidth.

Recently, with the rise of metamaterials that go beyond natural materials in many aspects, a lot of efforts have been devoted to the reexamination of the Brewster effect[7-17]. Generalized Brewster effect has been realized for enhanced transmission at selected angles[13,17]. Wide-angle Brewster effect has also been realized[14-16]. The bandwidth is relatively limited for resonant meta-structures[13-16], but it can be extended to an extremely wide spectrum from dc to THz and even higher frequencies for effective media such as the metallic gratings[8-11]. Despite these inspiring achievements, these generalized Brewster effects still face the same handicap as the traditional Brewster effect: unavoidable reflection at the presence of large absorption.

In this work, by designing anisotropic metamaterials, we reveal an anomalous Brewster effect (ABE) that allows independently tunable absorption and refraction in an ultra-broadband spectrum from dc to optical frequencies. No matter how large the absorption is, zero reflection can be maintained for p-polarization, which is impossible in previous traditional and generalized Brewster effects. Such an amazing effect is bestowed by anisotropic metamaterials designed under the guidance of the reciprocity principle. In this metamaterial, the anisotropy introduces extra degrees of freedoms to tune the absorption and refraction without affecting the Brewster condition. Therefore, loss-independent Brewster effect can be realized where the damping can be sufficiently large to achieve almost total absorption within a thickness of 1~2 wavelength. By performing proof-of-principle experiments in the microwave regime, we have successfully verified the theory. Our work provides a practical route to solve the long-standing dilemma between the Brewster effect and absorption.

The schematic graph of our idea is shown in Fig. 1(a). The anisotropic metamaterial is designed according to the reciprocity principle. The reciprocity principle requires that the response of a propagation channel is symmetric when the source and detection points are interchanged. For a wave incident and reflected on the surface of a reciprocal material[18], the switch between the incident angles of $\theta_i$ and $-\theta_i$ is equivalent to the exchange between the source and detection. Therefore, according to the reciprocity principle, the reflection coefficients are exactly the same for incident angles $\pm\theta_i$, i.e. $r(-\theta_i) = r(\theta_i)$ [18,19]. As a result, when the Brewster effect emerges, it is always simultaneously satisfied for $\pm\theta_i$. However, interestingly, upon the change between $\theta_i$ and $-\theta_i$, the absorption and transmission can be changed dramatically. And this makes the ABE possible.

The anisotropic metamaterial is designed such that a trivial solution of the Brewster effect (with zero loss and fixed angle of refraction) is obtained at $\theta_i > 0$ in an ultra-broad spectrum. At the same time, a nontrivial ABE can always be obtained at $\theta_i < 0$, where the absorption and refraction can be independently tuned by the extra parameters introduced by anisotropy. When the material is lossless, the angle of refraction can be tuned flexibly from positive to negative. When the material is lossy, ultra-broadband and controllable absorption can be achieved. Reflectionless Brewster absorbers can thus be realized by such non-resonant and nonmagnetic metamaterials, whose working bandwidth in principle can cover an extremely wide spectrum from dc to the optical regime. This result is remarkable as the bandwidth goes significantly beyond those of current techniques utilizing complex engineering of the material permeability and dispersions[5,6], where a tremendous amount of effort has been devoted. This huge bandwidth attributes to the inherent frequency independence of the Brewster effect.

**Results**

Without loss of generality, we consider a TM-polarized wave incident on a nonmagnetic tilted anisotropic medium (TAM) under the incident angle of $\theta_i$, as illustrated in Fig. 1(a). The background is a dielectric medium with isotropic permittivity $\varepsilon_b$. The optical axis (along with the permittivity component $\varepsilon_{x'}$) of the TAM has a tilt angle $\alpha$ with respect to the $x$ axis. For $\theta_i > 0$, we find that there are two trivial solutions for realizing the Brewster effect between the background and TAM which is independent of $\varepsilon_{y'}$. One solution for $\varepsilon_{x'} \neq \varepsilon_b$ requires a condition of $\theta_i = \pi/2 - \alpha$ ( $\alpha = \arctan\left(\sqrt{\varepsilon_b}/\sqrt{\varepsilon_{x'}}\right)$ ), which is exactly the condition of the traditional Brewster effect. The other solution for $\varepsilon_{x'} = \varepsilon_b$ requires a condition of $\theta_i = \alpha$ ( $\alpha$ is arbitrary), which is also easily understandable (see Supplementary Information). Intriguingly, in both cases, the waves incident at $\theta_i > 0$ cannot "see" $\varepsilon_{y'}$ as the electric field of the refracted beam $\mathbf{E}_t$ is parallel to the direction of $\varepsilon_{x'}$, i.e. $\mathbf{E}_t \parallel \varepsilon_{x'}$. This inspires us to utilize the new parameter $\varepsilon_{y'}$ as a control parameter without affecting the impedance matching condition. In the following, we provide a more comprehensive picture based on the equal-frequency contours (EFCs) of the metamaterial.

Figures 1(b) and 1(e) show the EFCs of the TAM with fixed $\varepsilon_{x'}$ and $\alpha$ for different $\varepsilon_{y'}$ under the conditions of $\varepsilon_{x'} \neq \varepsilon_b$ and $\varepsilon_{x'} = \varepsilon_b$, respectively. It is seen that in both figures, all the EFCs of the TAM cross a fixed point in the $k_y < 0$ space, which is marked as the point I (at $k_x = \sqrt{\varepsilon_{x'}} k_0 \sin\alpha$, $k_y = -\sqrt{\varepsilon_{x'}} k_0 \cos\alpha$, where $k_0$ is the wave number in free space). When the propagating state at the point I is excited in the TAM (by a proper incident angle $\theta_i > 0$), the electric field of the refracted wave $\mathbf{E}_t$ is polarized along the direction of $\varepsilon_{x'}$ (i.e. $\mathbf{E}_t \parallel \varepsilon_{x'}$), indicating that the waves cannot be influenced by $\varepsilon_{y'}$. Here, the wave impedance $Z$ is defined as the ratio of $E_x/H_z$, where $E_x$ and $H_z$ are, respectively, the $x$-component of electric fields and the $z$-component of magnetic fields at the interface (normal to $y$). Through some derivation, we obtain $|Z| = Z_0 \cos\alpha/\sqrt{\varepsilon_{x'}}$ at the point I for the TAM, which confirms that $|Z|$ is independent of $\varepsilon_{y'}$. Here, $Z_0$ is the impedance of vacuum (~377Ω). More importantly, the same impedance $|Z| = Z_0 \cos\alpha/\sqrt{\varepsilon_{x'}}$ is invariant along a vertical line of $k_x = \sqrt{\varepsilon_{x'}} k_0 \sin\alpha$ for various EFCs which goes through point I, when we vary $\varepsilon_{y'}$. This effect can be clearly seen in Figs. 1(b) and 1(e) where $|Z|$ is plotted in colors in the EFCs. Therefore, this vertical line is denoted as an equal-$|Z|$ line (black dashed lines) here. The analytical proof is summarized in and Supplementary Information.

Here, we would like to emphasize that despite that the two trivial solutions of the Brewster effect for $\theta_i > 0$ are both independent of $\varepsilon_{y'}$, they cannot provide the flexible manipulation of refraction or absorption yet as we mentioned in the introduction. This is because refracted waves cannot "see" $\varepsilon_{y'}$ and therefore, cannot be influenced by $\varepsilon_{y'}$ in any way. However, interesting things happen when considering the reciprocity principle. Now, we consider flipping the sign of the incident angle from $\theta_i > 0$ to $\theta_i < 0$ with the magnitude unchanged. According to the reciprocity principle, the Brewster effect is guaranteed and independent of $\varepsilon_{y'}$. Therefore, we have another equal-$|Z|$ line at $k_x = -\sqrt{\varepsilon_{x'}}k_0 \sin\alpha$, as marked by black dashed lines in Figs. 1(b) and 1(e). In the case of $\theta_i < 0$, the propagating states at point II are excited in the TAM, where the equal-$|Z|$ line and EFC intersect. Obviously, the EFC and the refraction angle both vary significantly with $\varepsilon_{y'}$. In this situation, the electric field of the refracted wave $\mathbf{E}_t$ is no long parallel to the direction of $\varepsilon_{x'}$, i.e. $\mathbf{E}_t \nparallel \varepsilon_{x'}$, bestowing tunable refractive behaviors by changing $\varepsilon_{y'}$, while the Brewster effect is guaranteed by reciprocity at the same time. When $\varepsilon_{y'}$ has an imaginary part, the absorption can be introduced. Therefore, we have obtained two nontrivial solutions of the ABE with tunable refraction and absorption.

In the following, we perform numerical simulations to prove the reciprocity-protected Brewster effect with tunable refraction in the background of air (i.e. $\varepsilon_b = 1$). The simulations were performed by the commercial finite-element-method software COMSOL Multi-physics. Here, we consider TM-polarized waves incident on the TAM. First, the TAM is chosen to be $\varepsilon_{x'} = 2$ and $\alpha = \arctan\left(\sqrt{\varepsilon_b}/\sqrt{\varepsilon_{x'}}\right) = 35.3°$. The reflectance on the air-TAM interface as a function of the $\theta_i$ and $\varepsilon_{y'}$ is plotted in Fig. 1(c). Clearly, near-zero reflection is observed around $\theta_i = \pm 54.7° = \pm(90° - \alpha)$, irrespective of $\varepsilon_{y'}$. We also notice blue lines related to omnidirectional Brewster effect, which can be understood through coordinate transformation[16,20]. In Fig. 1(d), we plot the distributions of magnetic fields (color) and group velocity (arrows) for the cases of $\theta_i = 54.7°$ (left), $\theta_i = -54.7°$ and $\varepsilon_{y'} = 3$ (middle), and $\theta_i = -54.7°$ and $\varepsilon_{y'} = -3$ (right), respectively. In all three cases, the reflection clearly disappears. By tuning $\varepsilon_{y'}$ from 3 to -3, the angle of refraction is evidently changed from positive to negative, which confirms the controllability of refraction by tuning $\varepsilon_{y'}$. Second, we consider a TAM with $\varepsilon_{x'} = 1$. In this situation, $\alpha$ can be arbitrary. Here, we take $\alpha = 30°$ as an example. The reflectance on the air-TAM interface is plotted in Fig. 1(f), which clearly shows $\varepsilon_{y'}$-independent near-zero reflection under $\theta_i = \pm 30°$, i.e. $\theta_i = \pm \alpha$. The reflection-less refraction for $\theta_i = \pm 30°$, as well as the controllable refraction for $\theta_i = -30°$ has also been well confirmed by simulations, as shown in Fig. 1(g).

It is worth noting that when $\varepsilon_{y'}$ approaches infinity, the EFCs of the TAM turn into two parallel lines with constant wave impedance $|Z| = Z_0 \cos\alpha / \sqrt{\varepsilon_{x'}}$, as are also shown as cyan color in Figs. 1(b) and 1(e). Such a case may be realized by using a tilted aluminum film array, which leads to ultra-broadband reflection-less negative refraction (see Supplementary Information).

As an important consequence of the above reciprocity-protected Brewster effect, we can further let $\varepsilon_{y'}$ be a complex number and introduce loss into the system. Such reciprocity consideration allows us to realize perfect-impedance-matched absorbing materials with unprecedented wide bandwidth, which is extremely difficult via other approaches, if not impossible. Such absorbers are denoted as reflectionless Brewster absorbers here. In the following, we demonstrate the design and realization of this ultra-broadband reflectionless Brewster absorbers by using tilted conductive film (CF) arrays.

The tilted CF array is embedded in a dielectric host of $\varepsilon_d$ under a tilt angle of $\alpha$, as illustrated by the upper inset graph of Fig. 2(a). The separation distance between two adjacent CFs is $a$, which is much larger than the thickness of the CFs $t$ (i.e. $a \gg t$), but smaller than the wavelength $\lambda$ in the dielectric host (i.e. $a < \lambda$). Under such circumstances, the tilted CF array can be approximately homogenized as an effective TAM (the lower inset) with $\varepsilon_{x',\text{eff}} = \varepsilon_d$ and $\varepsilon_{y',\text{eff}} = \varepsilon_d + i\gamma(\omega)$, where $\gamma(\omega) = (Z_0/R_s)/(k_0 a \cos\alpha)$ is a function of angular frequency $\omega$. Here, $R_s$ is the sheet resistance of the CFs. Figure 2(b) presents the equal-frequency surface of the effective TAM in the three-dimensional $k$ space composed of the $k_x$, $\text{Re}(k_y)$, and $\text{Im}(k_y)$ coordinates, which is obtained by adopting different values of $\gamma(\omega)$. The EFCs regarding the particular cases $\gamma(\omega) = 0$ and $\gamma(\omega) = \varepsilon_d$ are plotted as black and green lines, respectively. In the absence of material loss ($\gamma(\omega) = 0$), the EFC is a circle on the $k_x$-$\text{Re}(k_y)$ plane. When loss is introduced, the EFC separates into two curves extending into the $\text{Im}(k_y)$ direction. Interestingly, we find that the EFCs of the TAM with any $\gamma(\omega)$ always pass through a fixed point I located at the coordinates $k_x = \sqrt{\varepsilon_d} k_0 \sin\alpha$, $\text{Re}(k_y) = -\sqrt{\varepsilon_d} k_0 \cos\alpha$, and $\text{Im}(k_y) = 0$ (see the enlarged inset graph), indicating that it is independent of $\gamma(\omega)$. Similar to the lossless case, we have an equal-$|Z|$ surface at $k_x = \sqrt{\varepsilon_d} k_0 \sin\alpha$, which crosses the point I, and another equal-$|Z|$ surface at $k_x = -\sqrt{\varepsilon_d} k_0 \sin\alpha$, whose intersection with the equal-frequency surface turns into the ring II, as shown in Fig. 2(b). At the point I and on the ring II, the wave impedance is always a constant $|Z| = Z_0 \cos\alpha/\sqrt{\varepsilon_d}$. Interestingly, on the ring II, the induced $\text{Im}(k_y) \neq 0$ by $\gamma(\omega) \neq 0$ indicates the intriguing possibility to achieve large absorption without breaking the impedance matching condition.

We have also performed numerical simulations to verify the reciprocity-protected ABE with large absorption. Here, we consider TM-polarized waves incident on the TAM from an air background. First, the TAM is set as $\varepsilon_d = 2$ and $\alpha = 35.3°$. In Fig. 2(c), we plot the reflectance on the air-TAM interface as a function of $\theta_i$ and $\gamma(\omega)$, which clearly shows a regime of near-zero reflection around $\theta_i = \pm 54.7°$, irrespective of $\gamma(\omega)$. In Fig. 2(d), we plot the simulated magnetic field distributions for Gaussian beams incident on a TAM with $\gamma(\omega) = 3$ at $\theta_i = \pm 54.7°$. Clearly, for $\theta_i = 54.7°$ (the upper inset), the Gaussian beam transmits through the TAM, with almost no reflection or absorption. While for $\theta_i = -54.7°$ (the lower inset), the Gaussian beam is almost totally absorbed by the TAM, with no reflection or transmission. Second, the TAM is set as $\varepsilon_d = 1$. The reflectance on the air-TAM interface is plotted in Fig. 2(e), in which the TAM has $\alpha = 30°$. Clearly, near-zero reflection emerges around $\theta_i = \pm 30°$, irrespective of $\gamma(\omega)$. Interestingly, in this case, $\alpha$ is arbitrary and the Brewster effect is realized as long as $\theta_i = \pm \alpha$. This characteristic could be valuable for the absorption of the near field. For a demonstration, in Fig. 2(f), we demonstrate the simulated magnetic field distributions when a dipole source is placed above two inhomogeneous TAMs with variant $\alpha$ along the surface. The condition of $\theta_i = \alpha$ (upper inset) or $\theta_i = -\alpha$ (lower inset) is satisfied everywhere on the surface of the TAMs, where the $\gamma(\omega)$ is set as 2 and the orientation of the axis $\varepsilon_{y'}$ is denoted by arrows. Interestingly, in the case of $\theta_i = \alpha$, the radiation waves of the dipole source are almost totally transmitting through the TAM, with no reflection. While in the case of $\theta_i = -\alpha$, almost all the radiated waves from the dipole source are absorbed, with no reflection. This indicates a transition from total transparency to total absorption, simply by varying the orientation of the axis of the TAM. In other words, we first insert the parallel plates in a way that they do not perturb the fields as all these parallel plates act as waveguides without cut-off to allow unit transmittance. Then, all these conducting plates have the orientation flipped horizontally, allowing reciprocity-protected zero reflection while locally the E-fields have components parallel to the plates to generate huge absorption.

**Experimental observation of ultra-broadband reflectionless Brewster absorbers.** Since the reciprocity-protected ABE is irrespective of $\gamma(\omega)$, ultra-broadband reflection-less absorbers of electromagnetic waves can be realized. In the following, by using the tilted CF arrays, we experimentally verify this phenomenon in microwave frequencies. A sample of the CF array is fabricated, which consists of alternative polymethyl methacrylate (PMMA, relative permittivity ~2.6) and indium tin oxide (ITO) films (thickness~100μm, as CFs) with a separation distance of $a$=5mm. The sample has a tilt angle of $\alpha = 31.8°$, a thickness of 30mm, and a height of 200mm, as shown by the photo shown in Fig. 3(a). Ultra-broadband Brewster effect appears at $\theta_i = \pm 58.2°$ for such a TAM, while absorption only emerges at $\theta_i = -58.2°$. In Fig. 3(a), we plot the $|\text{Im}(k_y)|$ in the CF array (red dots) and the corresponding effective medium (black lines) as the function of the sheet resistance $R_s$ of the CFs at 10GHz. Here, $|\text{Im}(k_y)|$ determines the attenuation rate, and a larger $|\text{Im}(k_y)|$ implies a thinner thickness requirement for the absorber. In Fig. 3(a), the maximal $|\text{Im}(k_y)|$ is found to be around the optimal sheet resistance $\sim 0.35 Z_0 =132\Omega$ for the CF array, which coincides well with the effective medium calculation, indicating the validity of effective medium here. In the experiments, each piece of ITO film has a sheet resistance of ~370Ω. Therefore, we stacked two ITO films into one, so that the composite film has a sheet resistance of ~185Ω (blue dashed lines in Fig. 3(a)), and relatively good performance of absorption can be attained. Figures 3(b) and 3(c) show, respectively, the simulated reflectance and absorbance of the sample. In Fig. 3(b), near-zero reflection emerges around $\theta_i = \pm 58.2°$ for all frequencies below 16GHz. Above 16GHz, the non-reflection effect gradually deteriorates because the effective medium turns inaccurate in the short wavelength regime. Nevertheless, we note that the upper working frequency can be significantly raised, even to the optical frequency, by simply decreasing the scale of the structures, i.e. the separation distance. While the absorbance in Fig. 3(c) shows a distinct asymmetric behavior. Zero absorbance occurs around $\theta_i = 58.2°$ and high absorbance (>0.9) occurs around $\theta_i = -58.2°$ in the frequency regime from 7~15GHz. When the incident angle deviates away from the optimal angle $-58.2°$, the absorption is still quite high, showing some angular insensitivity in the absorption performance. Here, we note that the relatively low absorbance below 7GHz is mainly induced by transmission since the reflectance is almost zero. Since there is no reflection, and waves in the TAM decay exponentially with the propagation distance, the absorption at low frequencies can be easily increased through increasing the sample thickness.

The experimental results are shown in Fig. 3(d). In the experiment, an emitting horn antenna is placed ~0.8 meters away from the sample to generate the incident waves. We measured the near-field electric fields before and after the sample by using a probing antenna. The scanning areas are 700×150mm$^2$ each, located on the central plane of the sample. In the following, we demonstrate the measured electric field distribution at 7GHz, 10GHz and 15GHz (marked by stars in Figs. 3(b) and 3(c)). The upper (lower) inset graph of Fig. 3(d) presents the measured electric fields at $\theta_i = 58.2°$ ($\theta_i = -58.2°$). Clearly, there is no reflection in all cases, demonstrating the ultra-broadband ABE with absorption. Interestingly, near-total transmission is observed at $\theta_i = 58.2°$, while large absorption is clearly observed at $\theta_i = -58.2°$, for all three selected frequencies. In experiments, we evaluate the absorbance through integrating far-field power for all directions, finding absorbance as high as 0.95, 0.98, and 0.99 at 7GHz, 10GHz, and 15GHz, respectively. The absorbance at other frequencies in the measuring frequency range 7-15GHz is also quite high (see Supplementary Information).

In the second experiment, we verify the other nontrivial solution of the ultra-broadband ABE with absorption for the TAM designed with $\varepsilon_d = 1$. In this case, $\alpha$ is arbitrary and the Brewster effect is realized as long as $\theta_i = \pm\alpha$. The TAM sample is composed of alternative foam (relative permittivity ~1) and ITO films (thickness~100μm, as CFs) with a separation distance of $a$=10mm. The sample has a tilt angle of $\alpha = 30°$, a thickness of 30mm, and a height of 200mm, as shown by the photo shown in Fig. 4(a). In Fig. 4(a), we plot the $|\text{Im}(k_y)|$ in the sample (red dots) and the corresponding effective medium (black lines) as the function of the sheet resistance $R_s$ of the CFs at 10GHz. The maximal $|\text{Im}(k_y)|$ is observed around the optimal sheet resistance $Z_0\cos\alpha/(k_0 a)$=155Ω. Theoretically, the maximal value $|\text{Im}(k_y)|$ is derived to be $k_0\sin\alpha\tan\alpha$. In the experiment, we have exploited a composite film with $R_s$ ~185Ω consisting of two ITO films as adopted in the previous experiment (see blue dashed lines in Fig. 4(a)). Figures 4(b) and 5(c) show, respectively, the simulated reflectance and absorbance of the sample. Clearly, ultra-low reflection is achieved around $\theta_i = \pm 30°$ for all frequencies below 16GHz. Near-perfect absorption is achieved around $\theta_i = -30°$ in the frequency regime from 7~15GHz, while there is almost no absorption around $\theta_i = 30°$. Even when the incident angle deviates away from $-30°$, the absorption is still quite high. These results were also confirmed by experimental measurement. The measured electric field distributions are plotted in Fig. 4(d) for $\theta_i = 30°$ (upper inset) and $\theta_i = -30°$ (lower inset), respectively, at 7GHz, 10GHz and 15GHz. Indeed, near-total transmission is observed at $\theta_i = 30°$, while large absorption is clearly observed at $\theta_i = -30°$, for all three selected frequencies. The inset graphs of Fig. 4(d) show the measured far-field radiation patterns (green and red lines) of the sample. The black dashed lines denote the reference patterns in the absence of the sample. Clearly, ultra-broadband ABE with near-perfect absorption (>0.87 at 7GHz, >0.95 at 10GHz, >0.99 at 15GHz) is obtained.

There are several ways to further improve the absorption performance of the reflectionless Brewster absorbers demonstrated above. The simplest method is to increase the thickness of the samples since the metamaterial absorber is perfectly impedance-matched to free space. By adding a reflector behind the CF arrays, the angular-asymmetric performance of absorption would be changed into angular-symmetric large absorption, which may be important in practical applications (see Supplementary Information). In the case of $\varepsilon_d = \varepsilon_b$, the rotation angle $\alpha$ can be arbitrary. Interestingly, the maximal value $|\text{Im}(k_y)|$ is derived as $|\text{Im}(k_y)| = k_0 \sin\alpha \tan\alpha$, which tends to infinity as $\alpha$ goes to $90°$. This means that ultra-broadband perfect absorption can be realized even when $\alpha \to 90°$. Experimentally, we have also fabricated another sample with $\varepsilon_d = 1$ and $\alpha = 45°$, showing higher absorption (see Supplementary Information). Ultrathin perfect absorbers with extremely thin thickness can also be obtained based on the TAM with hyperbolic dispersions (see Supplementary Information).

**Discussion**

The hallmark advantage of the reflectionless Brewster absorbers is that they are based on inherently non-resonant metamaterials, and thus the bandwidth of impedance-matched absorption can cover the ultra-broad spectrum from dc to optical frequencies, which is far beyond those of other absorber techniques[5,6,21-36]. As a demonstration, in Fig. 5(a), we plot the reflection coefficent at the interface of free space and effective medium of the reflectionless Brewster absorber studied in Fig. 4 as the function of the incident angle, showing that zero reflection at $\theta_i = 30°$ (trivial case without absorption) and $\theta_i = -30°$ (nontrivial case with large absorption by the reflectionless Brewster absorber) from dc to GHz frequencies. There is no lower limit for the working frequency, while the upper limit is determined by the validity of the effective medium approximation. Through the reduction of the microstructure unit (e.g. the thickness and the separation distance of the CFs), the working frequency range of perfect-impedance-matched absorption can, in principle, be further widened to cover an unprecedented broad regime from dc to THz, infrared and even optical regimes (see Supplementary Information). Another fundamental advantage of the reflectionless Brewster absorbers lies in the tunability of the magnitude of material loss as well as the refractive behaviors, which offers significant flexibility.

Finally, we'd like to emphasize the difference between the traditional Brewster effect, the perfect electric conductor (PEC) grating structures[8-11] and the ABE proposed here. It has been observed that ultra-broadband Brewster effect can occur in dispersion-less effective media as well as PEC grating structures. However, when material loss is introduced, such Brewster effects will gradually break down, just like the traditional Brewster effect. The reflection can be dramatic when the loss is large. For verification, we consider a dielectric medium and a PEC grating structure exhibiting a Brewster angle at $75°$ in the absence of material loss for TM-polarized waves incident from free space. The period $d$ of the grating and the slit width $w$ satisfies $w/d = \cos 75°$. When we gradually increase the loss, dramatic reflection occurs at the air-dielectric and air-grating interfaces, as shown by the simulation results shown in Fig. 5(b). Here, $\text{Re}(\varepsilon)$ and $\text{Im}(\varepsilon)$ denote, respectively, the real and imaginary parts of the permittivity of the dielectric ($\text{Re}(\varepsilon) = \tan^2 75°$ to ensure a Brewster angle at $75°$), or the filling material in grating slits ($\text{Re}(\varepsilon)=1$). The inset graphs show the simulated magnetic-field distribution when $\text{Im}(\varepsilon)/\text{Re}(\varepsilon)=5$. In contrary, when the material loss is introduced to a TAM consisting of subwavelength tilted CF arrays in a dielectric host ($\varepsilon_d = \tan^2 75°$, $\alpha = -15°$, relative permittivity of CFs $\varepsilon = \text{Re}(\varepsilon)+i\,\text{Im}(\varepsilon)$), the reflection always remains almost zero irrespective of the magnitude of the loss, as shown in Fig. 5(b). Even when $\text{Im}(\varepsilon)/\text{Re}(\varepsilon)=500$, the reflection is still negligible, as seen from the magnetic-field distribution in the inset in Fig. 5(b).

In conclusion, in this work, we reveal an ABE for ultra-broadband reflectionless manipulation of waves, including tunable absorption and refraction. The ABE, as protected by the reciprocity principle, guarantees zero reflection for one polarization at a particular incident angle. At the same time, the refraction and absorption are flexibly tunable via the extra degrees of freedoms introduced by the anisotropy of metamaterials. The ABE bestows reflectionless Brewster absorbers with an unprecedented wide bandwidth of impedance-matched absorption. While conventional wisdom tells us that addition of loss will destroy the Brewster effect, we have demonstrated that the mechanism of reciprocity protection escapes from such a deficiency. This principle of ABE is universal for any reciprocal materials for general classical waves.

**Materials and methods**

**Simulations.** Numerical simulations in this work are performed using the commercial finite-element simulation software COMSOL Multiphysics. In order to simulate the TM-polarized plane wave under oblique incidence in Figs. 1d and 1g, the left and right boundaries are set as Floquet periodic boundaries, the top boundary is set as port to excite the incident wave. In Fig. 2d, the Gaussian beam is excited by the top port boundary. In Fig. 2f, the TAM is illuminated by an electric dipole source perpendicular to the TAM surface. The distance of the TAM surface equals the wavelength in air. In Figs. 2d and 2f, perfect matched layers are set in the bottom to absorb the transmission waves in the TAM.

**Experiments.** In near-field experiments, an emitting horn antenna is placed ~0.8 meter away from the sample to generate the incident waves. A probing antenna is used to probe the near-field electric fields on the central plane before and after the sample. The scanning rectangular area is of 700×150mm$^2$ (600×150mm$^2$) before and after the sample in Fig. 3 (Fig. 4). The probing antenna is mounted to a computer controlled translational stage. Both the probing antenna and the emitting horn antenna are connected to a network analyzer (KEYSIGHT PNA Network Analyzer N5224B) for data acquisition. Because of the selectivity of the probing antenna in measurement, only the electric fields perpendicular to the propagation direction are measured. Therefore, in the scan area before the sample, the measured electric fields come from the incident waves and a part of the possible reflection waves. In order to further confirm the zero reflection and near-perfect absorption in experiments, we have also measured the electric fields in the absence of samples and the far-field power radiation patterns.

In far-field experiments, an emitting horn antenna is placed 1m away from the sample to generate the incident waves. A receiving horn antenna placed at the same distance is used to measure the radiation pattern. The receiving horn antenna can be freely moved around the sample so that we could receive scattering signals in all directions. Both the emitting and receiving horn antennas are connected to a vector network analyzer (KEYSIGHT PNA Network Analyzer N5224B) for data acquisition. The power radiation pattern in the absence of sample is measured as a reference.

**Acknowledgements**

Y. L., R. P., and M. W. acknowledge support from the National Key R&D Program of China under Grant No. 2017YFA0303702, National Natural Science Foundation of China under Grant No. 61671314, 11974176, 11634005, 11974177, 61975078 and 11674155. J. Luo acknowledges support from the National Natural Science Foundation of China under Grant No. 11704271, Natural Science Foundation of Jiangsu Province under Grant No. BK20170326, and a Project Funded by the Priority Academic Program Development of Jiangsu Higher Education Institutions (PAPD). J. Li acknowledges support from the Research Grants Council of Hong Kong under Grant No. R6015-18.

**Conflict of interests**

The authors declare no competing interests.

**Contributions**

Y.L. and J.Luo conceived the idea. J.Luo formulated the theory and carried out the numerical simulations. J.Luo and J.Li established the reciprocity argument. H.C. carried out the experiments. All authors were involved in the discussion and analysis of data. Y.L., J.Luo and J.Li supervised the project and prepared the manuscript.

**Supplementary information** is available for this paper at denote the refracted beam determined by the EFCs of the TAM. [(c) and (f)] Calculated reflectance as functions of $\theta_i$ and $\varepsilon_{y'}$ for TM-polarized waves incident from free space onto the TAM with (c) $\varepsilon_{x'} = 2$ and $\alpha = 35.3°$, or (f) $\varepsilon_{x'} = 1$ and $\alpha = 30°$, respectively. [(d) and (g)] Simulated distributions of magnetic fields (color) and group velocity (arrows) for the TAM with (d) $\varepsilon_{x'} = 2$ and $\alpha = 35.3°$ under $\theta_i = 54.7°$ (left) or $\theta_i = -54.7°$ (middle and right), or (g) $\varepsilon_{x'} = 1$ and $\alpha = 30°$ under $\theta_i = 30°$ (left) or $\theta_i = -30°$ (middle and right), respectively. We set $\varepsilon_{y'} = 3$ (left and middle) or $\varepsilon_{x'} = -3$ (right).

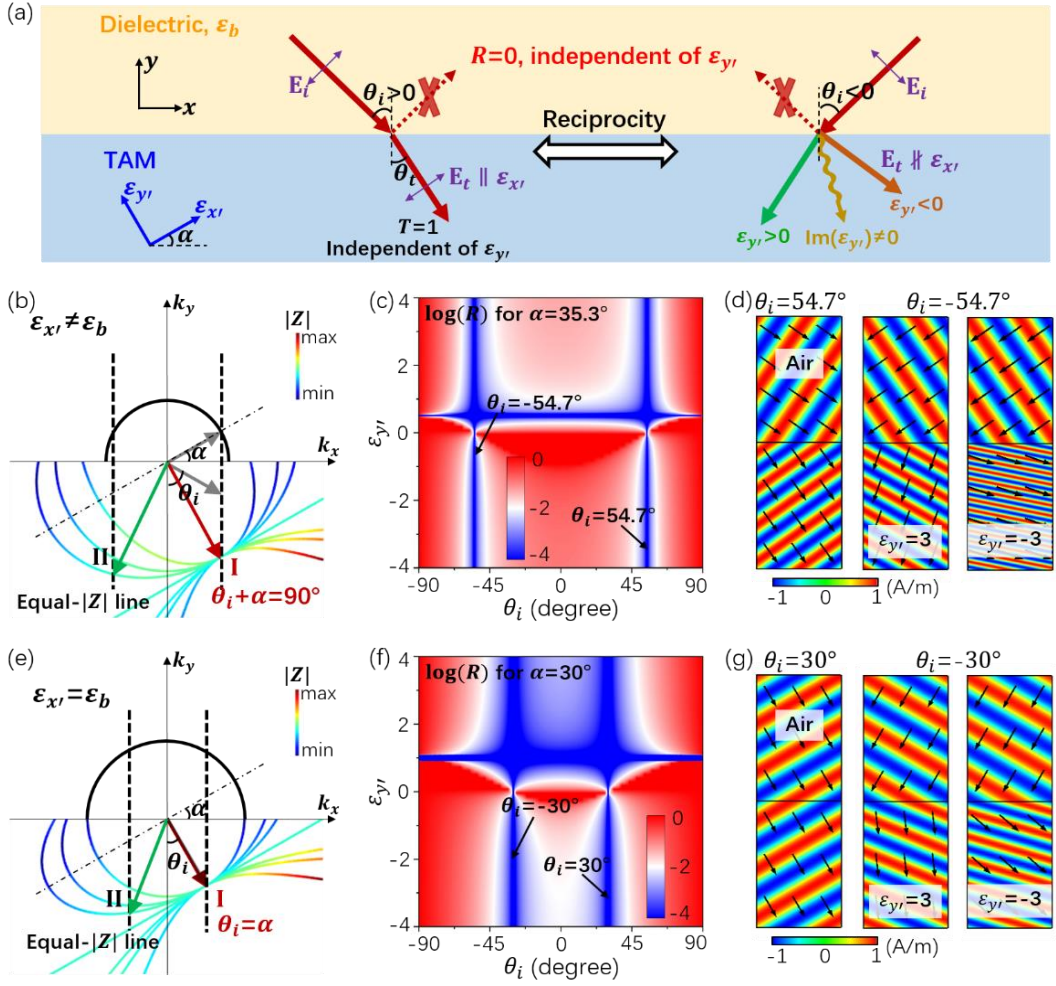

**Fig. 1 Simultaneous realization of $\varepsilon_{y'}$-independent Brewster effect and $\varepsilon_{y'}$-controlled transmission in TAMs.** (a) Illustration of the Brewster effect between a background dielectric $\varepsilon_b$ and a nonmagnetic TAM. Due to the protection of reciprocity, the wave impedance in the left case $\theta_i > 0$ is exactly the same as that in the right case $\theta_i < 0$. The transmission in the right (left) case is dependent (independent) on $\varepsilon_{y'}$. [(b) and (e)] The upper plane of EFCs of the background dielectric (black solid lines), and the lower plane of EFCs of the TAM when $\varepsilon_{x'}$ and $\alpha$ are fixed but $\varepsilon_{y'}$ is varied (color solid lines), regarding the cases of (b) $\varepsilon_{x'} \neq \varepsilon_b$ and (e) $\varepsilon_{x'} = \varepsilon_b$, respectively. The colors denote the magnitude of wave impedance $|Z|$. The black dashed vertical lines are the equal-$|Z|$ lines, which intersect with the EFCs at points I and II, respectively. The gray arrows denote the wave vectors of the incident and reflected beams determined by the EFCs of the background dielectric, and the red and green arrows

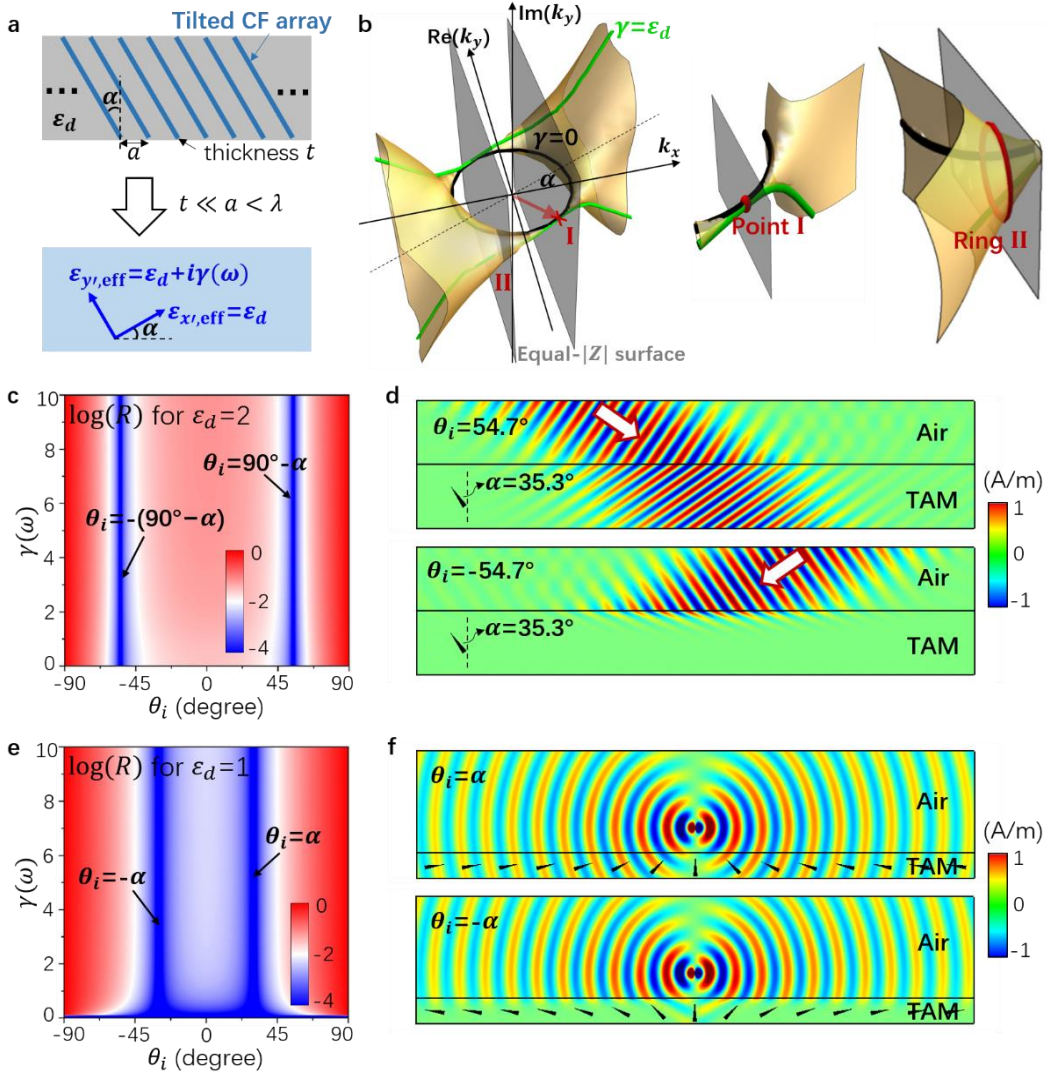

**Fig. 2 Ultra-broadband Brewster effect with controllable loss by using tilted CF arrays.** (a) Illustrations of a tilted CF array embedded in a dielectric of $\varepsilon_d$ (upper) as the effective TAM (lower). (b) The equal-frequency surface of the effective TAM that is formed by varying $\gamma(\omega)$. The black and green lines denote the EFCs regarding the cases of $\gamma(\omega)=0$ and $\gamma(\omega)=\varepsilon_d$, respectively. The gray surfaces denote the equal-$|Z|$ surfaces, which intersect with the equal-frequency surface at point I and the ring II (enlarge insets in the right). [(c) and (e)] Calculated reflectance as functions of the $\theta_i$ and $\gamma(\omega)$ for TM-polarized waves incident onto the effective TAM with (c) $\varepsilon_d=2$ and $\alpha=35.3°$, or (e) $\varepsilon_d=1$ and $\alpha=30°$, respectively. (d) Simulated magnetic field-distributions when a TM-polarized Gaussian beam is incident onto the effective TAM with $\varepsilon_d=2$, $\alpha=35.3°$ and $\gamma(\omega)=3$ for $\theta_i=54.7°$

(upper) or $\theta_i = -54.7°$ (lower). (f) Simulated magnetic field-distributions for a TM-polarized dipole source above an inhomogeneous TAM with fixed $\varepsilon_d = 1$ and $\gamma(\omega) = 3$ but varied $\alpha$. The $\alpha$ is adjusted along the surface so that the condition of $\theta_i = \alpha$ (upper) or $\theta_i = -\alpha$ (lower) is satisfied everywhere. The arrows in (d) and (f) denote the orientation of the axis $\varepsilon_{y'}$.

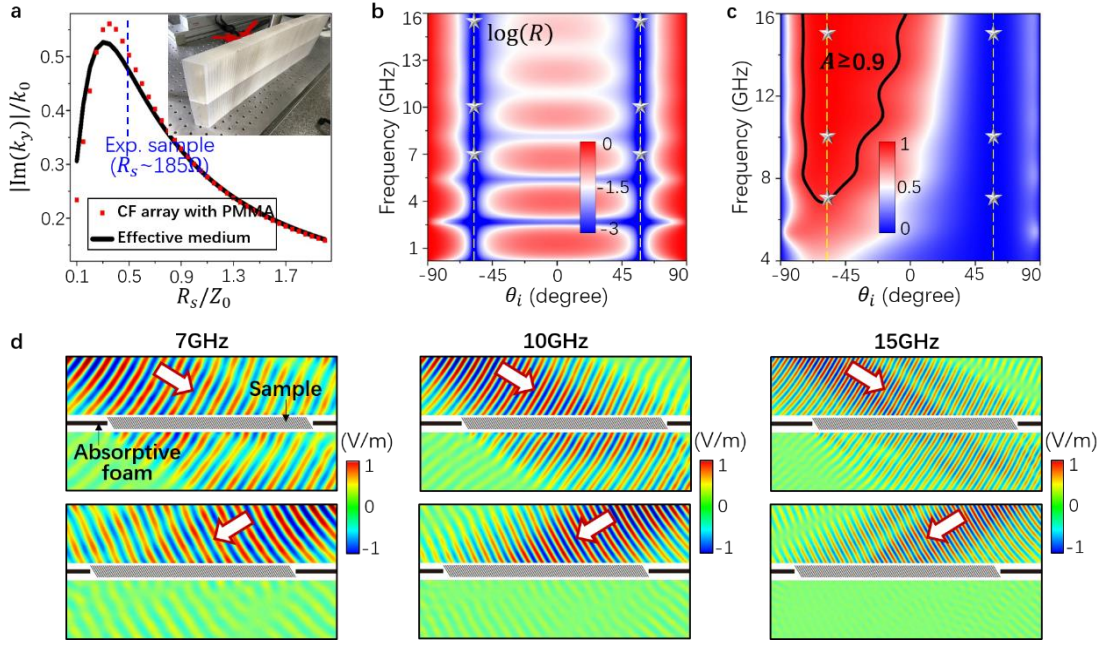

**Fig. 3 Experimental observation of ultra-broadband reflectionless Brewster absorbers in the case of $\varepsilon_d \neq \varepsilon_b$.** (a) The $|\text{Im}(k_y)|$ in the CF array (red dots) and the corresponding effective medium (black solid lines) as the function of the $R_s$ at 10GHz. The inset shows the picture of the fabricated sample consisting of alternative PMMA and ITO films $\alpha = 31.8°$. The thickness and height of the sample are 30mm and 200mm, respectively. [(b) and (c)] Simulated (b) reflectance and (c) absorbance on the fabricated sample as functions of the $\theta_i$ and the working frequency, respectively. The stars denote the measured cases in experiments. (d) Measured electric field-distributions under $\theta_i = 58.2°$ (upper) and $\theta_i = -58.2°$ (lower) at 7GHz (left), 10GHz (middle) and 15GHz (right). The black thick lines near the sample denote the absorptive foam.

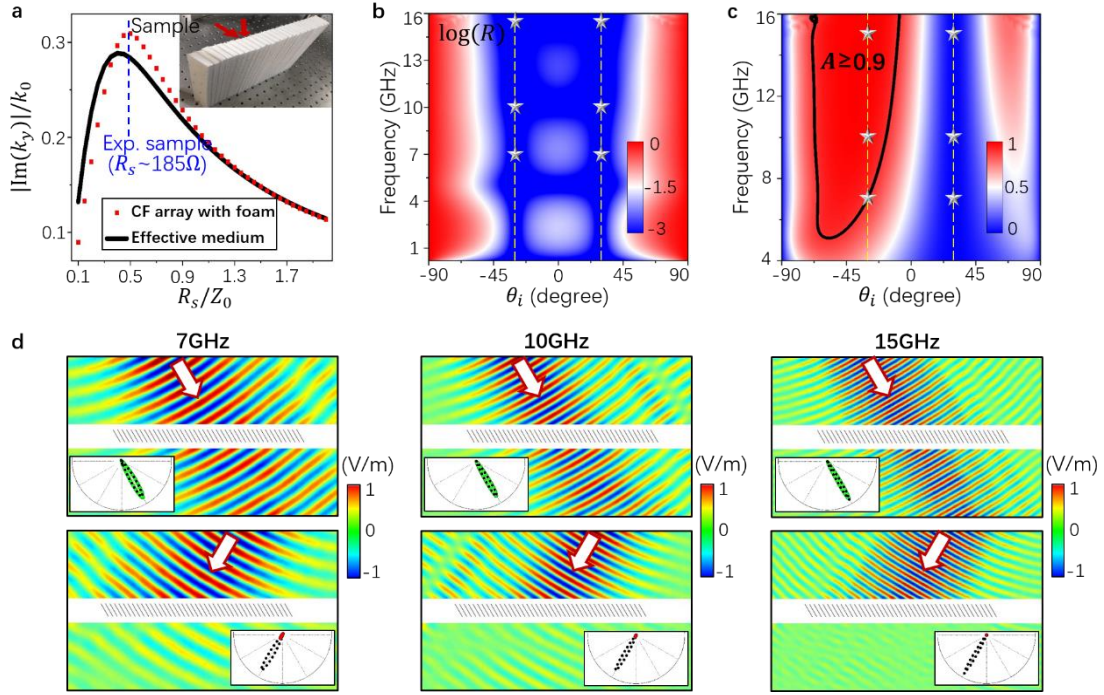

**Fig. 4 Experimental observation of ultra-broadband reflectionless Brewster absorbers in the case of** $\varepsilon_d = \varepsilon_b$. (a) The $|\text{Im}(k_y)|$ in the CF array (red dots) and the corresponding effective medium (black solid lines) as the function of the $R_s$ at 10GHz. The inset shows the picture of the fabricated sample consisting of an alternative foam and ITO films. [(b) and (c)] Simulated reflectance and absorbance on the fabricated sample as functions of the $\theta_i$ and the working frequency, respectively. The stars denote the measured cases in experiments. (d) Measured electric field-distributions under $\theta_i = 30°$ (upper) and $\theta_i = -30°$ (lower) at 7GHz (left), 10GHz (middle) and 15GHz (right). The insets show the measured far-field radiation patterns with (green and red solid lines) or without (black dashed lines) samples.

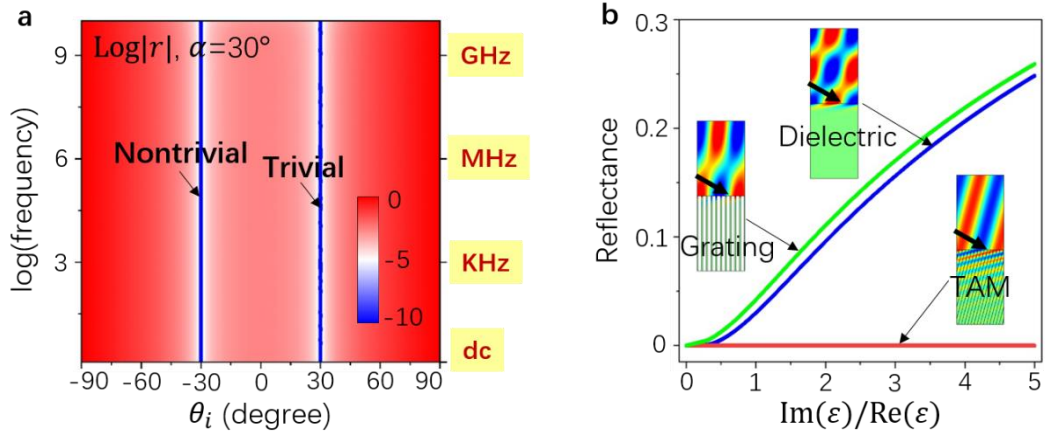

**Fig. 5 The ultra-broad bandwidth of ABE and the dependence of reflection on absorption.** (a) Reflection coefficient as the function of the $\theta_i$ at the interface of free space and the effective medium model of the CF array in Fig. 4. The nontrivial solution corresponds to the ABE that enables flexible large absorption and tunable angle of refraction, while the trivial solution is similar to the traditional Brewster effect, without absorption and tunable refraction. (b) Reflectance as the function of material loss at the interfaces of air-dielectric (blue lines), air-PEC-grating (green lines) and air-TAM (red lines) under $\theta_i = 75°$. The TAM is composed of subwavelength tilted CF array in a dielectric host with $\varepsilon_d = \tan^2 75°$ and $\alpha = -15°$. $\mathrm{Re}(\varepsilon)$ and $\mathrm{Im}(\varepsilon)$ denote the real and imaginary parts of the permittivity of the dielectric, or the filling material in grating slits, or the CF of the TAM, respectively. The insets show the simulated magnetic-field distributions when $\mathrm{Im}(\varepsilon)/\mathrm{Re}(\varepsilon) = 5$ for the dielectric and grating models, $\mathrm{Im}(\varepsilon)/\mathrm{Re}(\varepsilon) = 500$ for the TAM model.

# Supplemental Materials

# Ultra-broadband reflectionless Brewster absorber protected by reciprocity

Jie Luo[1*#], Hongchen Chu[2#], Ruwen Peng[2], Mu Wang[2], Jensen Li[3,4*], Yun Lai[2*]

[1]School of Physical Science and Technology, Soochow University, Suzhou 215006, China

[2]National Laboratory of Solid State Microstructures, School of Physics, and Collaborative Innovation Center of Advanced Microstructures, Nanjing University, Nanjing 210093, China

[3]Department of Physics, The Hong Kong University of Science and Technology, Clear Water Bay, Hong Kong, China

[4]William Mong Institute of Nano Science and Technology, the Hong Kong University of Science and Technology, Clear Water Bay, Kowloon, Hong Kong, China

*Corresponding authors: Yun Lai (laiyun@nju.edu.cn); Jie Luo (luojie@suda.edu.cn); Jensen Li (jensenli@ust.hk)

#These authors contributed equally: Jie Luo, Hongchen Chu

1. **Derivations of wave impedance and conditions of ABE**
2. **Complementary discussions of the lossless TAM**
3. **Ultra-broadband reflection-less positive and negative refraction, and experimental verification**
4. **Loss-induced breakdown of the Brewster effect in isotropic dielectrics**
5. **ABE and perfect absorption in general TAM with material loss**
6. **Extraordinary ultrathin perfect absorbers by TAM with hyperbolic dispersions**
7. **Effective medium model of tilted CF array and ABE from dc to the GHz regime**
8. **Optimal sheet resistance analysis, details of experimental samples and further experimental results**
9. **Experimental setup and measurement methods**
10. **Improvement of absorption by using reflectors**

1. **Derivations of wave impedance and conditions of ABE**

We consider a transverse-magnetic (TM, magnetic field along the $z$ direction) polarized wave propagating on the $x$-$y$ plane. The wave is incident from an isotropic dielectric (relative permittivity $\varepsilon_b$) onto a nonmagnetic tilted anisotropic medium (TAM). Such a TAM is characterized by uniaxial relative permittivity components $\varepsilon_{x'}$ and $\varepsilon_{y'}$, together with a rotation angle $\alpha$ (see Fig. 1 in the Main Text). The reflection coefficient (with respect to the magnetic field) at the dielectric-TAM interface can be derived as,

$$r^{TM} = \frac{(\bar{\varepsilon}_{xx}\bar{\varepsilon}_{yy} - \bar{\varepsilon}_{xy}\bar{\varepsilon}_{yx})k_{b,y} - \varepsilon_b(\bar{\varepsilon}_{xy}k_x + \bar{\varepsilon}_{yy}k_y)}{(\bar{\varepsilon}_{xx}\bar{\varepsilon}_{yy} - \bar{\varepsilon}_{xy}\bar{\varepsilon}_{yx})k_{b,y} + \varepsilon_b(\bar{\varepsilon}_{xy}k_x + \bar{\varepsilon}_{yy}k_y)}, \quad (S1)$$

where $\bar{\varepsilon}_{xy}$ and $\bar{\varepsilon}_{yy}$ are the components of the relative permittivity tensor $\bar{\varepsilon}$ in the un-rotated $x$-$y$ coordinate, i.e. $\bar{\varepsilon} = \begin{pmatrix} \bar{\varepsilon}_{xx} & \bar{\varepsilon}_{xy} \\ \bar{\varepsilon}_{yx} & \bar{\varepsilon}_{yy} \end{pmatrix} = \begin{pmatrix} \varepsilon_{x'}\cos^2\alpha + \varepsilon_{y'}\sin^2\alpha & (\varepsilon_{x'} - \varepsilon_{y'})\sin\alpha\cos\alpha \\ (\varepsilon_{x'} - \varepsilon_{y'})\sin\alpha\cos\alpha & \varepsilon_{x'}\sin^2\alpha + \varepsilon_{y'}\cos^2\alpha \end{pmatrix}$.

The $k_{b,y}$ and $k_x$ are connected by the dispersion of the isotropic dielectric, i.e., $k_x^2 + k_{b,y}^2 = \varepsilon_b k_0^2$ with $k_0$ being the wave number in free space. For a certain incident angle $\theta_i$, we have $k_x = \sqrt{\varepsilon_b}k_0\sin\theta_i$ and $k_{b,y}^f = -\sqrt{\varepsilon_b}k_0\cos\theta_i$ for forward propagating waves (or $k_{b,y}^b = \sqrt{\varepsilon_b}k_0\cos\theta_i$ for backward propagating waves). The $k_y$ and $k_x$ are connected by the dispersion of the TAM, i.e.

$$k_x^2\bar{\varepsilon}_{xx} + k_y^2\bar{\varepsilon}_{yy} + k_xk_y(\bar{\varepsilon}_{xy} + \bar{\varepsilon}_{yx}) = (\bar{\varepsilon}_{xx}\bar{\varepsilon}_{yy} - \bar{\varepsilon}_{xy}\bar{\varepsilon}_{yx})k_0^2 \quad \text{or}$$

$$(k_x^2\varepsilon_{x'} + k_y^2\varepsilon_{y'})\cos^2\alpha + (k_y^2\varepsilon_{x'} + k_x^2\varepsilon_{y'})\sin^2\alpha + k_xk_y(\varepsilon_{x'} - \varepsilon_{y'})\sin 2\alpha = \varepsilon_{x'}\varepsilon_{y'}k_0^2.$$

According to Eq. (S1), we find that the **condition of the non-reflection or perfect impedance matching (PIM)** is,

$$\frac{k_{b,y}}{\varepsilon_b} = \frac{\bar{\varepsilon}_{xy}k_x + \bar{\varepsilon}_{yy}k_y}{\bar{\varepsilon}_{xx}\bar{\varepsilon}_{yy} - \bar{\varepsilon}_{xy}\bar{\varepsilon}_{yx}}. \quad (S2)$$

The left term in Eq. (S2) is proportional to the wave impedance $Z_b$ of the isotropic dielectric, i.e. $Z_b = \frac{E_x}{H_z} = -\frac{k_{b,y}}{\varepsilon_b}\frac{Z_0}{k_0}$. Here, $E_x$ and $H_z$ are, respectively, the $x$-component of electric field and $z$-component of magnetic field in the isotropic dielectric. $Z_0$ is the vacuum impedance (~377Ω). The right term in Eq. (S2) is proportional to the wave impedance of the TAM:

$$Z_{TAM} = -\frac{\overline{\varepsilon}_{xy} k_x + \overline{\varepsilon}_{yy} k_y}{\overline{\varepsilon}_{xx} \overline{\varepsilon}_{yy} - \overline{\varepsilon}_{xy} \overline{\varepsilon}_{yx}} \frac{Z_0}{k_0} . \qquad (S3)$$

According to the dispersion of the TAM, we find out two solutions of $k_y$ for a fixed $k_x$, that is,

$$k_y^f = \frac{-\sqrt{\varepsilon_{x'}}\sqrt{\varepsilon_{y'}}\sqrt{k_0^2 (\varepsilon_{x'} \sin^2 \alpha + \varepsilon_{y'} \cos^2 \alpha) - k_x^2} + k_x \sin 2\alpha (\varepsilon_{y'} - \varepsilon_{x'})/2}{\varepsilon_{y'} \cos^2 \alpha + \varepsilon_{x'} \sin^2 \alpha} \quad \text{and}$$

$$k_y^b = \frac{\sqrt{\varepsilon_{x'}}\sqrt{\varepsilon_{y'}}\sqrt{k_0^2 (\varepsilon_{x'} \sin^2 \alpha + \varepsilon_{y'} \cos^2 \alpha) - k_x^2} + k_x \sin 2\alpha (\varepsilon_{y'} - \varepsilon_{x'})/2}{\varepsilon_{y'} \cos^2 \alpha + \varepsilon_{x'} \sin^2 \alpha},$$ which are related to the forward

and backward propagating waves, respectively. By substituting the two solutions into the wave impedance in Eq. (S3), we find that,

$$Z_{TAM}(k_x, k_y^f) = -Z_{TAM}(k_x, k_y^b) = \frac{\sqrt{k_0^2 (\varepsilon_{x'} \sin^2 \alpha + \varepsilon_{y'} \cos^2 \alpha) - k_x^2}}{\sqrt{\varepsilon_{x'}} \sqrt{\varepsilon_{y'}}} \frac{Z_0}{k_0} . \qquad (S4)$$

Since the wave impedance in Eq. (S4) is an even function of $k_x$, we have $Z_{TAM}(k_x, k_y^f) = Z_{TAM}(-k_x, k_y^f)$ and $Z_{TAM}(k_x, k_y^b) = Z_{TAM}(-k_x, k_y^b)$. Therefore, **for any fixed $k_x$, the absolute value of the wave impedance is a constant**, that is,

$$\left| Z_{TAM}(\pm k_x, k_y^f) \right| = \left| Z_{TAM}(\pm k_x, k_y^b) \right| . \qquad (S5)$$

Now, by taking dispersions of the isotropic dielectric and the TAM into account, the PIM condition in Eq. (S2) can be rewritten as,

$$\varepsilon_{x'} \varepsilon_{y'} k_0^2 \cos^2 \theta_i + \varepsilon_b \left[ k_x^2 - k_0^2 (\varepsilon_{x'} \sin^2 \alpha + \varepsilon_{y'} \cos^2 \alpha) \right] = 0 . \qquad (S6)$$

We assume that the PIM condition is $\varepsilon_{y'}$-independent. The $\varepsilon_{y'}$-independence means that the first derivative of the left term in Eq. (S6) versus $\varepsilon_{y'}$ is zero, thus we have

$$\varepsilon_{x'} \cos^2 \theta_i = \varepsilon_b \cos^2 \alpha . \qquad (S7)$$

Alternatively, we can obtain the condition of $\varepsilon_{y'}$-independence in a physical way. When transmission waves in the TAM propagate along the axis of $\varepsilon_{y'}$, the electric field will polarize along the axis of $\varepsilon_{x'}$. In this situation, the refraction angle is $\alpha$, which is irrelevant to $\varepsilon_{y'}$. Considering the refraction law, we get the relationship between the $\theta_i$ and $\alpha$ as,

$$\sqrt{\varepsilon_b} \sin \theta_i = \sqrt{\varepsilon_{x'}} \sin \alpha . \qquad (S8)$$

By substituting Eq. (S7) or Eq. (S8) into Eq. (S6), we get the same result, that is,

$$\left(\varepsilon_b - \varepsilon_{x'}\tan^2\alpha\right)\left(\varepsilon_b - \varepsilon_{x'}\right) = 0. \qquad (S9)$$

**Equation (S9) is the condition of anomalous Brewster effect (ABE), i.e. $\varepsilon_{y'}$-independent PIM.**

Equation (S9) indicates two classes of solutions to the $\varepsilon_{y'}$-independent PIM:

1) **One is $\varepsilon_b - \varepsilon_{x'}\tan^2\alpha = 0$ (or $\alpha = \arctan\left(\sqrt{\varepsilon_b}/\sqrt{\varepsilon_{x'}}\right)$) when $\varepsilon_b \neq \varepsilon_{x'}$. Combined with Eq. (S7) or Eq. (S8), we obtain $\theta_i = \pm(\pi/2 - \alpha)$.**

   When $\theta_i = \pi/2 - \alpha$, the PIM is similar to the classical Brewster angle effect. In this case, the transmission waves in the TAM propagate along the axis of $\varepsilon_{y'}$, which are also $\varepsilon_{y'}$-independent. However, when $\theta_i = -(\pi/2 - \alpha)$, the transmission waves will rely on $\varepsilon_{y'}$ although the PIM is still $\varepsilon_{y'}$-independent. **The same PIM under $\theta_i = \pi/2 - \alpha$ and $\theta_i = -(\pi/2 - \alpha)$ attributes to the reciprocity.**

2) **Another one is $\varepsilon_b = \varepsilon_{x'}$, which has no restriction on the value of $\alpha$. Based on Eq. (S7) or Eq. (S8), we obtain $\theta_i = \pm\alpha$.**

   When $\theta_i = \alpha$, we have trivial PIM. In this case, the transmission waves in the TAM propagate along the axis of $\varepsilon_{y'}$, which are also $\varepsilon_{y'}$-independent. However, when $\theta_i = -\alpha$, the transmission waves will rely on $\varepsilon_{y'}$ although the PIM is still $\varepsilon_{y'}$-independent. **The same PIM under $\theta_i = \alpha$ and $\theta_i = -\alpha$ attributes to the reciprocity.**

The above results reveal that for forward propagating waves, we have the same wave impedance when $k_x = \pm\sqrt{\varepsilon_b}k_0\cos\theta_i = \pm\sqrt{\varepsilon_{x'}}k_0\cos\alpha$, which is $\varepsilon_{y'}$-independent. Combined with Eq. (S5), we further obtain,

$$\left|Z_{TAM}\left(\pm\sqrt{\varepsilon_{x'}}k_0\sin\alpha, k_y^f\right)\right| = \left|Z_{TAM}\left(\pm\sqrt{\varepsilon_{x'}}k_0\sin\alpha, k_y^b\right)\right| = \frac{\cos\alpha}{\sqrt{\varepsilon_{x'}}}Z_0 \quad (\varepsilon_{y'}\text{-independent}) \qquad (S10)$$

Equation (10) indicates **two equal-$|Z_{TAM}|$ lines of $k_x = \pm\sqrt{\varepsilon_{x'}}k_0\cos\alpha$ for the TAM with varied** $\varepsilon_{y'}$, as shown in Fig. 1 in the Main Text.

## 2. Complementary discussions of the lossless TAM

First, we present more detailed discussions on the lossless TAM with $\varepsilon_b \neq \varepsilon_{x'}$. In order to show the relationship between the incident angle $\theta_i$ and the rotation angle $\alpha$ in the realization of $\varepsilon_{y'}$-independent PIM, we take the TAM with $\varepsilon_{x'} = 2$ as an example.

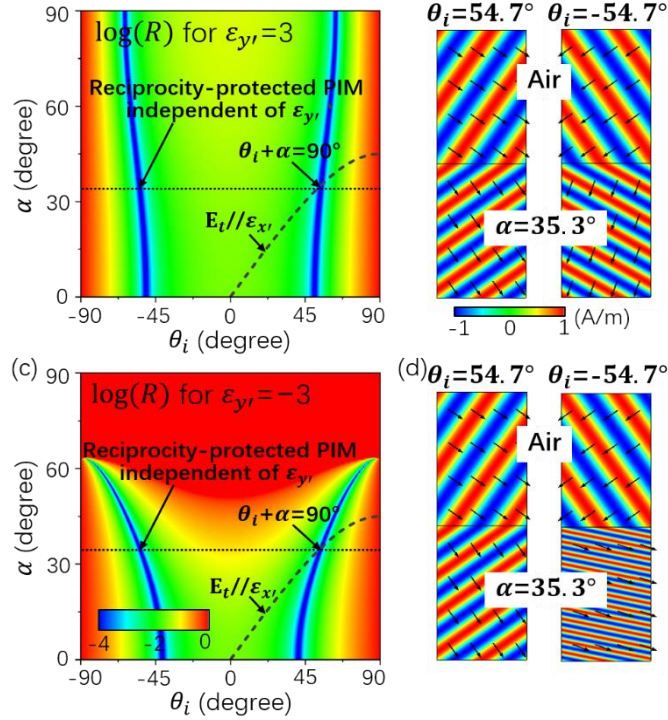

**Figure S1. The relationship between the incident angle and rotation angle in the case of $\varepsilon_b \neq \varepsilon_{x'}$.** [(a) and (c)] Reflectance at the air-TAM ($\varepsilon_{x'} = 2$) interface as functions of the $\theta_i$ and $\alpha$. The $\varepsilon_{y'}$ is chosen as (a) 3, (c) -3. The incident waves are of TM polarizations. [(b) and (d)] Distributions of magnetic fields (color) and group velocity (arrows) when a TM polarized wave is incident from air onto the TAM ($\varepsilon_{x'} = 2$, $\alpha = 35.3°$) with (b) $\varepsilon_{y'} = 3$, (d) $\varepsilon_{y'} = -3$ under $\theta_i = 54.7°$ (left) and $\theta_i = -54.7°$ (right).

Figures S1(a) and S1(c) plot the reflectance for TM-polarized waves incident from air onto the TAM with $\varepsilon_{y'} = 3$ and $\varepsilon_{y'} = -3$, respectively. The dark blue lines with zero reflectance satisfy Eq. (S6), where the TAM is impedance matched with air. Now, we consider **transmission waves propagating along the axis of $\varepsilon_{y'}$ in the TAM (the electric field $E_t$ is parallel to the axis of $\varepsilon_{x'}$), so that the PIM is $\varepsilon_{y'}$-independent**. In this situation, the relationship between the $\theta_i$ and $\alpha$ is determined by Eq. (S8). In Figs. S1(a) and S1(c), we plot the Eq. (S8) as black dashed lines, that intersect the impedance-matched lines (Eq. (S6)) in the $\theta_i > 0$ space. Although the $\varepsilon_{y'}$ is different, the intersection point in Figs. S1(a) and S1(c) is the same, i.e. $\alpha = \arctan\left(\sqrt{\varepsilon_b}/\sqrt{\varepsilon_{x'}}\right) \approx 35.3°$ and $\theta_i = 90° - \alpha = 54.7°$, which is related to the $\varepsilon_{y'}$-independent PIM. Interestingly, reciprocity leads to another point of $\varepsilon_{y'}$-independent PIM at $\alpha = \arctan\left(\sqrt{\varepsilon_b}/\sqrt{\varepsilon_{x'}}\right) \approx 35.3°$ and $\theta_i = -(90° - \alpha) = -54.7°$, as shown in Figs. S1(a) and S1(c). Moreover, we simulate the distributions of magnetic fields (color) and group velocity (arrows) under $\theta_i = 54.7°$ (left) and $\theta_i = -54.7°$ (right) in Figs. S1(b) and S1(d). When $\theta_i = 54.7°$, we can see the same wave phenomena, i.e. zero reflection and the same refraction angle, despite of the different $\varepsilon_{y'}$. When $\theta_i = -54.7°$, the reflection is still zero due to the protection of reciprocity. However, it is interesting to see that the varied $\varepsilon_{y'}$ can tune the transmission waves in the TAM, including the refraction angle, group velocity and propagation phase.

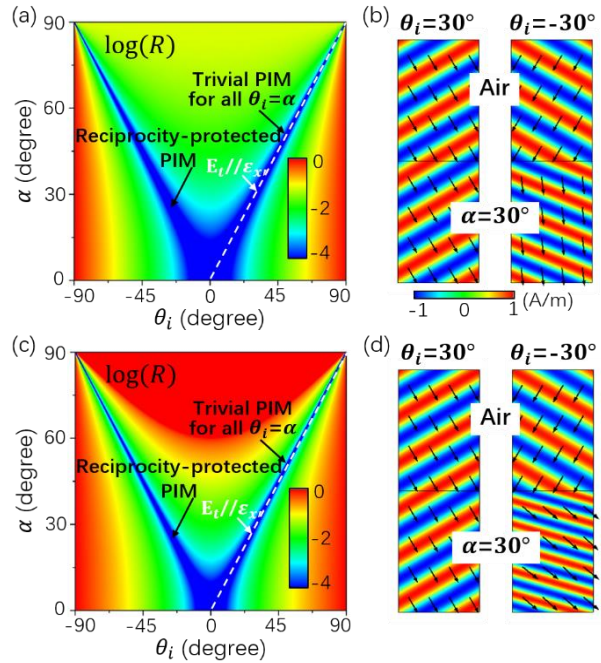

**Figure S2. The relationship between the incident angle and rotation angle in the case of $\varepsilon_b = \varepsilon_{x'}$.** [(a) and (c)] Reflectance at the air-TAM ($\varepsilon_{x'} = 1$) interface as functions of the $\theta_i$ and $\alpha$ when (a) $\varepsilon_{y'} = 3$, (c) $\varepsilon_{y'} = -3$. The incident waves are of TM polarizations. [(b) and (d)] Distributions of magnetic fields (color) and group velocity (arrows) when a TM polarized wave is incident from air onto the TAM ($\varepsilon_{x'} = 1$, $\alpha = 30°$) with (b) $\varepsilon_{y'} = 3$, (d) $\varepsilon_{y'} = -3$ under $\theta_i = 30°$ (left) and $\theta_i = -30°$ (right).

Second, we discuss more about the lossless TAM with $\varepsilon_b = \varepsilon_{x'}$. In this case, the condition of ABE in Eq. (S6) is simplified to,

$$(\varepsilon_b - \varepsilon_{y'})(\sin^2 \theta_i - \sin^2 \alpha) = 0, \qquad (S11)$$

which indicates the PIM for all $\theta_i = \pm \alpha$ when $\varepsilon_b \neq \varepsilon_{y'}$.

Without loss of generality, we consider that $\varepsilon_b = \varepsilon_{x'} = 1$. In Figs. S2(a) and S2(c), the TAM with $\varepsilon_{y'} = 3$ and $\varepsilon_{y'} = -3$ are studied as examples. The reflectance at the air-TAM interface is plotted as functions of the $\theta_i$ and $\alpha$. The dark blue lines of $\theta_i = \pm \alpha$ denote the condition of PIM (i.e. Eq. (S6) or Eq. (S11)). Furthermore, we plot the condition of $\varepsilon_{y'}$-independence (i.e. Eq. (S8)) as white dashed lines, which overlap with the dark blue line in the right side. This indicates that for all $\theta_i = \alpha$, we have the $\varepsilon_{y'}$-dependent PIM, and there is no restriction on the value of the $\alpha$. Considering the reciprocity, we have the same $\varepsilon_{y'}$-dependent PIM for all $\theta_i = -\alpha$. For verification, we take the TAM with $\alpha = 30°$ as an example. Figures S2(b) and S2(d) show the distributions magnetic fields (color) and group velocity (arrows) under $\theta_i = 30°$ (left) and $\theta_i = -30°$ (right), showing the $\varepsilon_{y'}$-independent zero reflection under $\theta_i = \pm 30°$, and the $\varepsilon_{y'}$-controlled transmission waves under $\theta_i = -30°$.

## 3. Ultra-broadband reflection-less positive and negative refraction, and experimental verification

In this section, we show that the $\varepsilon_{y'}$-independent PIM and $\varepsilon_{y'}$-controlled flexible refraction can be simultaneously obtained. In this way, we can realize both reflection-less positive and negative refraction. Figure S3(a) presents the EFCs of air (black dashed lines) and the TAM ($\varepsilon_{x'}=1$) with varied $\varepsilon_{y'}$ (blue solid lines). We know that when TM-polarized waves are incident from air onto the TAM under $\theta_i = \pm\alpha$, we have the $\varepsilon_{y'}$-independent PIM. The green arrows in Fig. S3(a) denote wave vectors under $\theta_i = \pm\alpha$. Considering the conservation of wave-vector component parallel to the air-TAM surface, we can find out the direction of group velocity $\mathbf{v}_g$ of transmission waves in the TAM, which is normal to the EFCs (red arrows). It is seen from Fig. S3(a) that the direction of $\mathbf{v}_g$ is $\varepsilon_{y'}$-independent under $\theta_i = \alpha$ (see the point I). However, interestingly, under $\theta_i = -\alpha$, the direction of $\mathbf{v}_g$ changes as the variation of $\varepsilon_{y'}$, and both reflection-less positive and negative refraction can be obtained (see the point II).

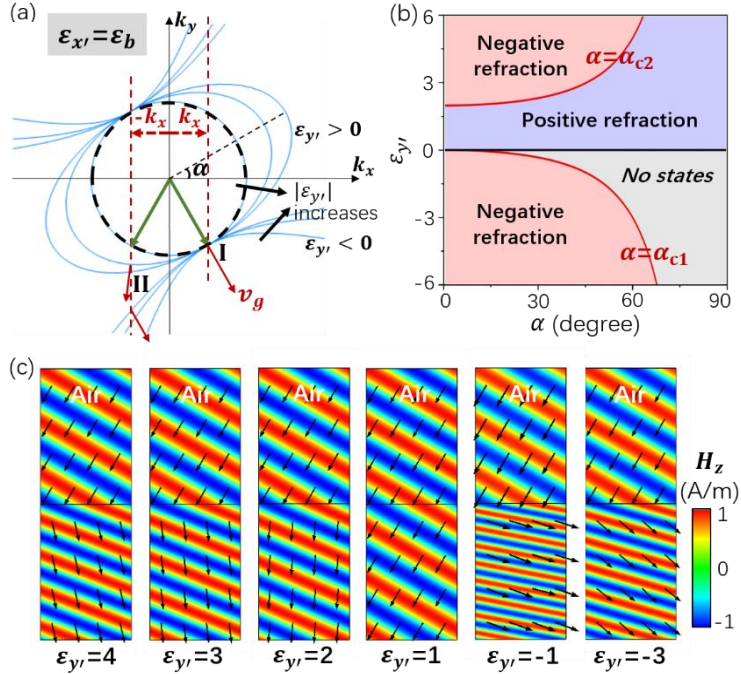

**Figure S3. Simultaneous realization of $\varepsilon_{y'}$-independent PIM and $\varepsilon_{y'}$-controlled flexible refraction.** (a) EFCs of air (black dashed lines) and the TAM ($\varepsilon_{x'}=1$) with varied $\varepsilon_{y'}$ (blue solid lines). The red arrows denote directions of group velocity of transmission waves in the TAM. (b) Phase diagram of reflection-less refraction in the $\alpha$-$\varepsilon_{y'}$ space under $\theta_i = -\alpha$. (c) Simulated distributions of magnetic

fields (color) and group velocity (arrows) for TM-polarized waves incident from air onto the TAM ($\varepsilon_{x'}=1$, $\alpha=30°$) with different $\varepsilon_{y'}$ under $\theta_i=-30°$.

Figure S3(b) shows the phase diagram of reflection-less positive and negative refraction in the $\alpha$-$\varepsilon_{y'}$ space under $\theta_i=-\alpha$. It is seen that when $\varepsilon_{y'}<0$, we generally have reflection-less negative refraction unless $\alpha>\alpha_{c1}$. Here $\alpha_{c1}$ is a critical angle determined by

$$\varepsilon_{y'}=-\varepsilon_{x'}\tan^2\alpha_{c1}. \qquad (S12)$$

When $\alpha=\alpha_{c1}$, one of the asymptotes of hyperbolic EFC of the TAM will overlap with the $k_y$ axis. Once $\alpha>\alpha_{c1}$, there will be no states supporting the forward propagating waves. Physically, it is not strictly correct, because wave vectors cannot be infinitely large and material losses are inevitable in practical hyperbolic media. As long as a small amount of material loss is considered, there will exist states supporting the forward propagating waves when $\alpha>\alpha_{c1}$.

On the other hand, when $\varepsilon_{y'}>0$, phase transition from reflection-less positive refraction to reflection-less negative refraction, or vice versa, will occur at the critical rotation angle $\alpha_{c2}$:

$$\varepsilon_{y'}=\frac{1}{2}\varepsilon_{x'}\left(3+\cos 2\alpha_{c2}\right)\sec^2\alpha_{c2}, \qquad (S13)$$

which is related to zero slope of the TAM EFC at the point II.

For demonstrations, we simulate the wave propagation from air onto the TAM with $\varepsilon_{x'}=1$ and $\alpha=30°$. Figure S3(c) displays the distributions of magnetic fields (color) and group velocity (arrows) under $\theta_i=-30°$ when $\varepsilon_{y'}$ is chosen as 4, 3, 2, 1, -1 and -3 for examples. Apparently, there are no reflections in all examples. Moreover, we see that the direction of group velocity of transmission waves in the TAM can be efficiently manipulated by $\varepsilon_{y'}$, thus demonstrating **the simultaneous realization of the $\varepsilon_{y'}$-independent PIM and $\varepsilon_{y'}$-controlled reflection-less positive and negative refraction**.

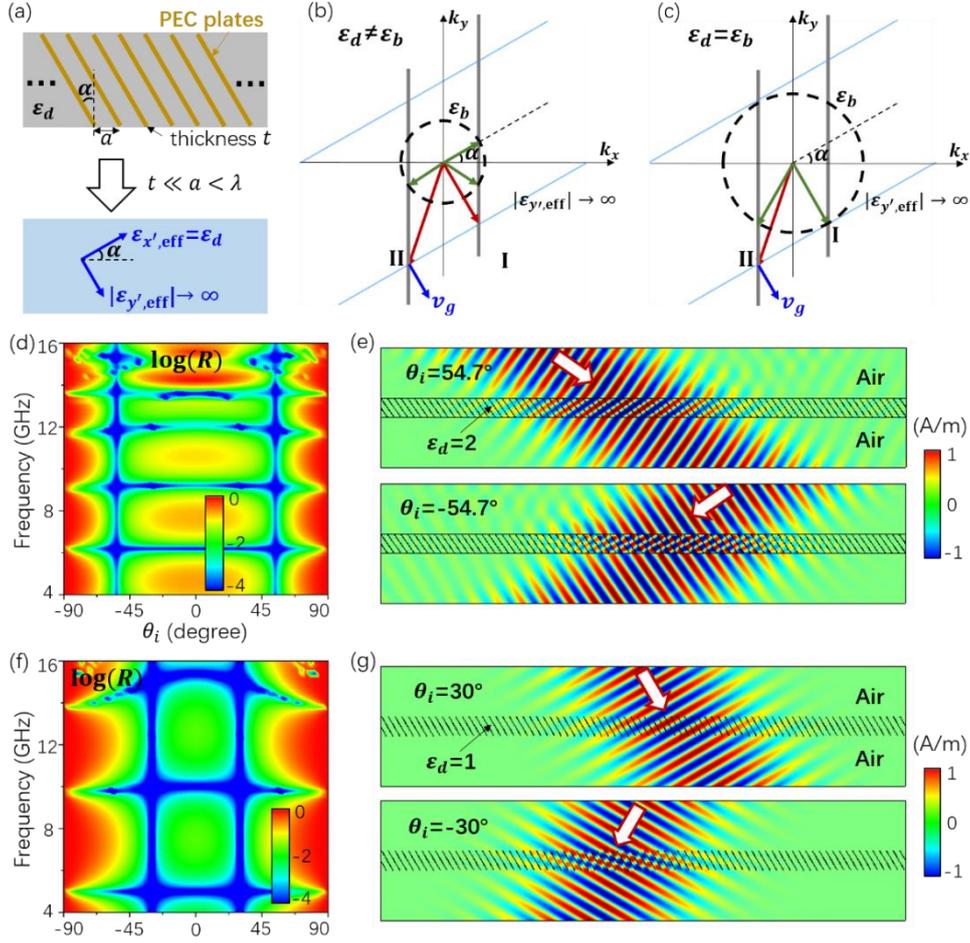

**Figure S4. Theoretical investigation of ultra-broadband reflection-less negative refraction by using tilted PEC plate arrays.** (a) Illustrations of a tilted PEC plate array in an isotropic dielectric of $\varepsilon_d$ (upper) and the corresponding effective TAM (lower). [(b) and (c)] The EFCs of a dielectric of $\varepsilon_b$ (dashed lines) and the effective TAM (solid lines) for the cases of (b) $\varepsilon_d \neq \varepsilon_b$ and (c) $\varepsilon_d = \varepsilon_b$. The green and red arrows denote the incident/reflection and transmission beams, respectively. The blue arrows show the direction of group velocity of the transmission waves in the TAM. [(d) and (f)] Reflectance as functions of the $\theta_i$ and working frequency when a TM-polarized wave is incident from air onto the tilted PEC plate array with (d) $\varepsilon_d = 2$ and $\alpha = 35.3°$, (f) $\varepsilon_d = 1$ and $\alpha = 30°$. [(e) and (g)] Simulated magnetic field-distributions at 10GHz for the cases of (e) $\varepsilon_d = 2$ and $\alpha = 35.3°$ under $\theta_i = \pm 54.7°$, (g) $\varepsilon_d = 1$ and $\alpha = 30°$ under $\theta_i = \pm 30°$.

Next, we demonstrate the ultra-broadband reflection-less negative refraction by using tilted perfect electric conductor (PEC) plates in both simulations and microwave experiments. The model we studied is illustrated by the upper inset in Fig. S4(a). Ultrathin PEC plates with a rotation angle of $\alpha$ are periodically aligned in an isotropic dielectric (relative permittivity $\varepsilon_d$). The separation distance between two adjacent PEC plates is $a$, which is much larger than the thickness of the PEC plates $t$ (i.e. $a \gg t$), but smaller than the wavelength $\lambda$ in the dielectric (i.e. $a < \lambda$). Under this circumstance, we can approximately homogenize the tilted PEC plate array as an effective TAM with

$$\varepsilon_{x',\text{eff}} \approx \varepsilon_d \quad \text{and} \quad |\varepsilon_{y',\text{eff}}| \to \infty. \tag{S14}$$

In this situation, the wave impedance of the effective TAM (i.e. Eq. (S4)) turns to be,

$$Z_{TAM} \approx \frac{\cos\alpha}{\sqrt{\varepsilon_d}} Z_0, \tag{S15}$$

which indicates a **frequency-independent wave impedance**. Based on this unique feature, ultra-broadband reflection-less negative reflection can be realized, as we shall demonstrate as follows.

Figure S4(b) shows the EFCs of the isotropic dielectric (dashed lines) and the effective TAM with $|\varepsilon_{y',\text{eff}}| \to \infty$ (solid lines) for the case of $\varepsilon_d \neq \varepsilon_b$. It is seen that when $\theta_i = \pi/2 - \alpha$ and $\alpha = \arctan\sqrt{\varepsilon_b/\varepsilon_d}$, the TAM operates at the point I, and the reflection beam (green arrows) is perpendicular to the refraction beam (red arrows). This would lead to the PIM similar to Brewster effect. We note that **such PIM is frequency-independent, because both wave impedances of the dielectric and the effective TAM are frequency-independent**. Considering the reciprocity principle, the frequency-independent PIM reserves under $\theta_i = -(\pi/2 - \alpha)$ (i.e. the TAM operates at the point II). Interestingly, in this case, we have negative refraction, as illustrated by the blue arrows in Fig. S4(b). Thus, ultra-broadband reflection-less negative refraction can be attained. Similarly, in the case of $\varepsilon_d = \varepsilon_b$, due to the frequency-independent PIM under $\theta_i = \alpha$ and the reciprocity, we can also get the frequency-independent PIM and negative refraction under $\theta_i = -\alpha$, as illustrated in Fig. S4(c).

For verification, we first study the case of $\varepsilon_d \neq \varepsilon_b$ by assuming a tilted PEC plate array ($a$=10mm, $\alpha$=35.3°) embedded in an isotropic dielectric of $\varepsilon_d = 2$. The array has a thickness of $d$=30mm, and is placed in the background of air (i.e. $\varepsilon_b = 1$). Figure S4(d) presents the reflectance as functions of the $\theta_i$ and working frequency under the illumination of a TM-polarized wave, showing frequency-independent zero reflection under $\theta_i = \pm 54.7°$. In Fig. S4(e), the simulated magnetic field-distributions under $\theta_i = 54.7°$ (upper) and $\theta_i = -54.7°$ (lower) at 10GHz clearly demonstrate the zero reflection and negative refraction at $\theta_i = -54.7°$. Second, we study the case of $\varepsilon_d = \varepsilon_b = 1$. The tilted PEC plate array is characterized by $a$=10mm, $d$=30mm and $\alpha$=30°. The calculated reflectance in Fig. S4(f) shows the frequency-independent zero reflection at $\theta_i = \pm 30°$. Moreover, the simulated magnetic field-distributions at 10GHz in Fig. S4(g) show obvious negative refraction under $\theta_i = -30°$. **Apparently, these results demonstrate the ultra-broadband reflection-less negative refraction by using a tilted PEC plate array.**

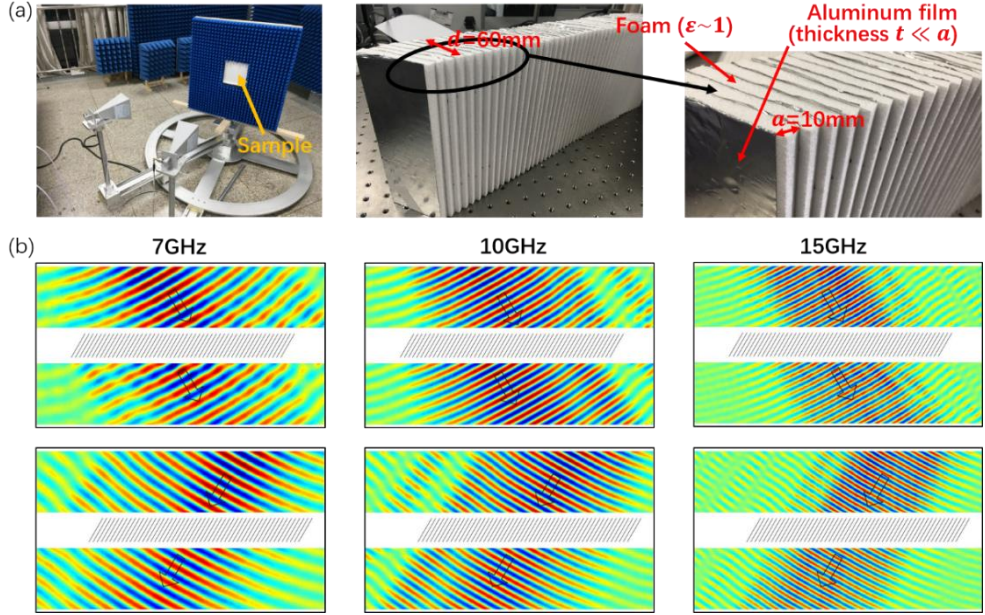

**Figure S5. Experimental observation of ultra-broadband reflection-less negative refraction.** (a) Photographs of the experimental setup and fabricated sample composed of alternative foam and aluminum films with $\alpha =30°$, $a =10$mm and $d =60$mm. (b) Measured electric fields under $\theta_i = 30°$ (upper) and $\theta_i = -30°$ (lower) at 7GHz (left), 10GHz (middle) and 15GHz (right).

We have also performed microwave experiments to demonstrate such ultra-broadband reflection-less negative refraction. Figure S5(a) presents the photographs of the experimental setup and fabricated sample. The sample is composed of alternative foam (relative permittivity ~1) and aluminum films (near PEC films) with $\alpha =30°$, $a =10$mm and $d =60$mm. The thickness of the aluminum films is ultrathin, much smaller than $a$. Figure S5(b) displays the measured near-field electric fields under $\theta_i = 30°$ (upper) and $\theta_i = -30°$ (lower) at 7GHz (left), 10GHz (middle) and 15GHz (right). The details of the experimental setup and measurement methods are discussed in the following Section 8. **The experimental results clearly confirm the ultra-broadband PIM under $\theta_i = \pm\alpha$, as well as ultra-broadband reflection-less negative refraction under $\theta_i = -\alpha$.**

## 4. Loss-induced breakdown of the Brewster effect in isotropic dielectrics

As a comparison of loss-independent Brewster effect in TAM, here we show that the Brewster effect breaks down immediately when loss is introduced to isotropic dielectrics. Figures S6(a) and 6(b) plot reflectance $\log(R)$ as functions of incident angle $\theta_i$ and $\gamma(\omega)$ for TM-polarized waves incident onto isotropic dielectric with $\varepsilon = 2 + \gamma(\omega)$ and $\varepsilon = 1 + \gamma(\omega)$, respectively. **Clearly, the Brewster effect can no longer be maintained and dramatic reflection occurs as long as the loss is introduced.**

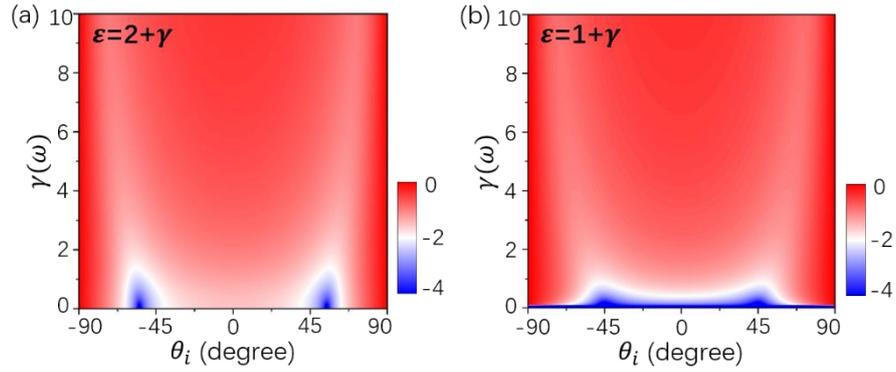

**Figure S6. Loss-induced breakdown of the Brewster effect in isotropic dielectrics.** Calculated reflectance $\log(R)$ as functions of incident angle $\theta_i$ and $\gamma(\omega)$ for TM-polarized waves incident onto isotropic dielectric with (a) $\varepsilon = 2 + \gamma(\omega)$, (b) $\varepsilon = 1 + \gamma(\omega)$.

## 5. ABE and perfect absorption in general TAM with material loss

In the Main Text, we have demonstrated the ABE and perfect absorption with tilted CF arrays, whose effective parameters satisfy $\varepsilon_{x'} = \text{Re}(\varepsilon_{y'})$. In this section, we discuss the ABE and perfect absorption in general TAM with $\varepsilon_{x'} \neq \text{Re}(\varepsilon_{y'})$.

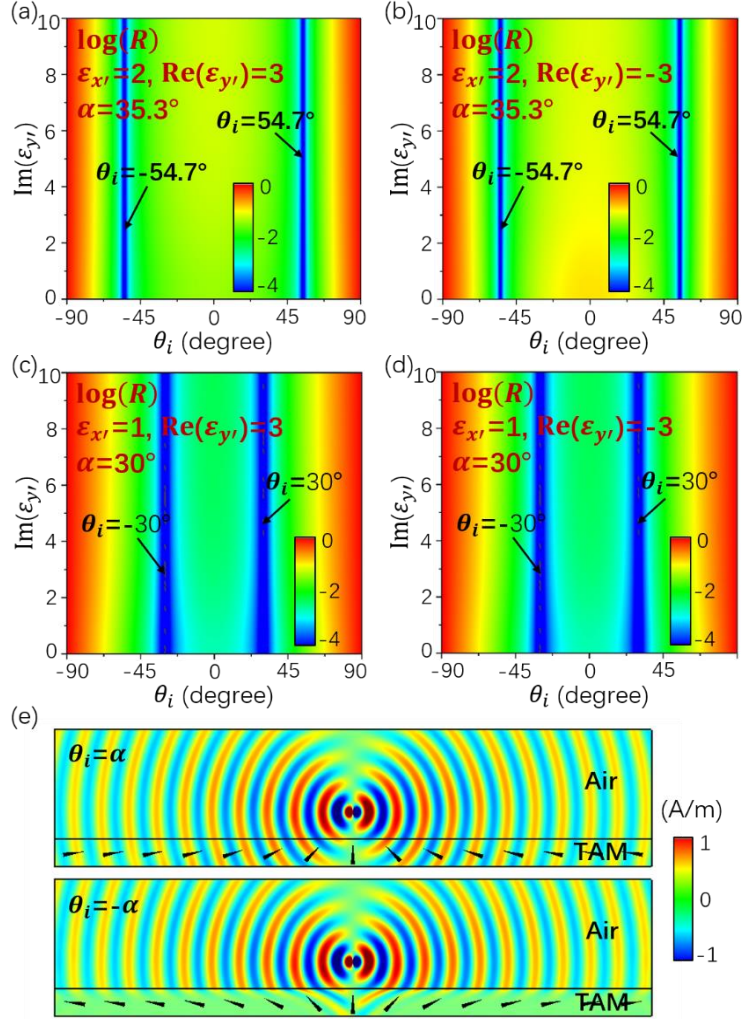

**Figure S7.** ABE and perfect absorption in general TAM with $\varepsilon_{x'} \neq \text{Re}(\varepsilon_{y'})$. [(a)-(d)] Reflectance of TM-polarized waves incident from air onto the TAM with (a) $\varepsilon_{x'} = 2$, $\text{Re}(\varepsilon_{y'}) = 3$, $\alpha = 35.3°$, (b) $\varepsilon_{x'} = 2$, $\text{Re}(\varepsilon_{y'}) = -3$, $\alpha = 35.3°$, (c) $\varepsilon_{x'} = 1$, $\text{Re}(\varepsilon_{y'}) = 3$, $\alpha = 30°$, (d) $\varepsilon_{x'} = 1$, $\text{Re}(\varepsilon_{y'}) = -3$, $\alpha = 30°$. (e) Simulated magnetic field-distributions when a TM-polarized dipole source is placed above an inhomogeneous TAM slab with fixed $\varepsilon_{x'} = 1$, $\varepsilon_{y'} = 3 + 3.33i$ and thickness $d = 30$mm at 10GHz. The orientation of $\varepsilon_{y'}$ of the TAM (black arrows) is engineered, so that the condition of $\theta_i = \alpha$ (upper)

or $\theta_i = -\alpha$ (lower) is satisfied everywhere.

Here, we take the TAM with $\varepsilon_{x'} = 2$, $\text{Re}(\varepsilon_{y'}) = \pm 3$, $\alpha = 35.3°$ in Figs. S7(a) and S7(b), and the TAM with $\varepsilon_{x'} = 1$, $\text{Re}(\varepsilon_{y'}) = \pm 3$, $\alpha = 30°$ in Figs. S7(c) and S7(d) as examples. Figures S7(a)-S7(d) show the reflectance as functions of the $\theta_i$ and $\text{Im}(\varepsilon_{y'})$ when TM-polarized waves incident from air onto the TAM. Apparently, $\text{Im}(\varepsilon_{y'})$-independent zero reflection is seen at $\theta_i = \pm 54.7°$ (or $\theta_i = \pm 30°$) for the TAM with $\varepsilon_{x'} = 2$ (or $\varepsilon_{x'} = 1$), demonstrating the ABE, i.e. $\varepsilon_{y'}$-independent PIM, in the existence of material loss.

Furthermore, in Fig. S7(e), we perform simulations by illuminating an inhomogeneous TAM slab ($\varepsilon_{x'} = 1$, $\varepsilon_{y'} = 3 + 3.33i$, thickness $d = 30$mm) with a TM-polarized dipole source at 10GHz. The orientation of $\varepsilon_{y'}$ of the TAM (black arrows) is engineered so that the condition of $\theta_i = \alpha$ (upper) or $\theta_i = -\alpha$ (lower) is satisfied everywhere. It is seen that in both cases of $\theta_i = \pm \alpha$, there are no reflection waves, demonstrating the omnidirectional PIM. Interestingly, when $\theta_i = -\alpha$, rapid attenuation of transmission waves in the TAM is observed, demonstrating the omnidirectional perfect absorption.

**The above results clearly demonstrate that PIM is completely independent of both the real and imaginary parts of $\varepsilon_{y'}$, and the perfect absorption can be realized in general TAM with $\varepsilon_{x'} \neq \text{Re}(\varepsilon_{y'})$.**

## 6. Extraordinary ultrathin perfect absorbers by TAM with hyperbolic dispersions

In this section, we first study the attenuation rate of transmission waves in TAM with material losses, and then demonstrate an extraordinary kind of ultrathin perfect absorbers by TAM with hyperbolic dispersions.

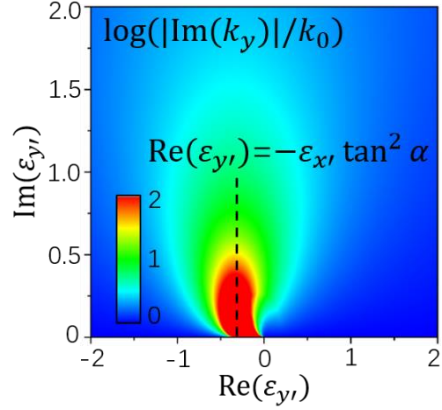

**Figure S8. Attenuation rate of transmission waves in TAM with material losses.** $\log\left(\left|\mathrm{Im}(k_y)\right|/k_0\right)$ as functions of the $\mathrm{Re}(\varepsilon_{y'})$ and $\mathrm{Im}(\varepsilon_{y'})$. $k_y$ is the $y$-component of wave vector of transmission waves in the TAM ($\varepsilon_{x'}=1$, $\alpha=30°$).

In Fig. S8, we take the TAM with $\varepsilon_{x'}=1$ and $\alpha=30°$ as an example. $\log\left(\left|\mathrm{Im}(k_y)\right|/k_0\right)$ of transmission waves in the TAM is calculated as functions of the $\mathrm{Re}(\varepsilon_{y'})$ and $\mathrm{Im}(\varepsilon_{y'})$, showing extraordinary large $\mathrm{Im}(k_y)$ at

$$\mathrm{Re}(\varepsilon_{y'}) = -\varepsilon_{x'}\tan^2\alpha. \tag{S16}$$

Equation (S16) indicates that $\varepsilon_{x'}$ and $\mathrm{Re}(\varepsilon_{y'})$ have opposite signs, that is, the TAM possesses a hyperbolic dispersion in the absence of material losses. Compared with Eq. (S12), we find that **the extraordinary large $\mathrm{Im}(k_y)$ occurs when one of the asymptotes of hyperbolic EFC overlaps with the $k_y$ axis. Intriguingly, when $\mathrm{Im}(\varepsilon_{y'})\approx 0$, infinitely large $\mathrm{Im}(k_y)$ can be obtained, revealing a unique kind of ultrathin perfect absorbers**. Moreover, it is seen that **the increase of $\mathrm{Im}(\varepsilon_{y'})$ will lead to the decrease of $\mathrm{Im}(k_y)$, showing an extraordinary absorption behavior**.

For verification, Fig. S9(a) displays magnetic field-distributions when a TM-polarized wave is incident from air onto the TAM with $\varepsilon_{x'} = 1$, $\varepsilon_{y'} = -3 + 0.1i$ and $\alpha = 60°$, showing non-absorption under $\theta_i = 60°$ (left) and perfect absorption within an ultrathin thickness under $\theta_i = -60°$ (right). For comparison, we increase the $\mathrm{Im}(\varepsilon_{y'})$, so that $\varepsilon_{y'} = -3 + 10i$ in Fig. S9(b). The zero absorption under $\theta_i = 60°$ (left) and perfect absorption under $\theta_i = -60°$ (right) are not changed. But the transmission waves in the TAM decay much more slowly than those in the TAM with $\varepsilon_{y'} = -3 + 0.1i$.

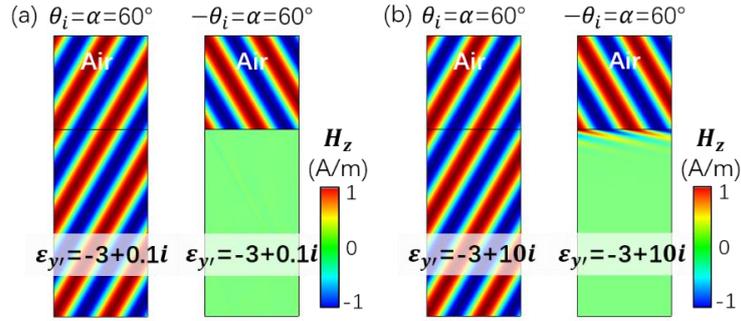

**Figure S9. Extraordinary perfect absorption by TAM with hyperbolic dispersions.** [(a) and (b)] Magnetic field-distributions when a TM-polarized wave is incident from air onto the TAM ($\varepsilon_{x'} = 1$, $\alpha = 60°$) with (a) $\varepsilon_{y'} = -3 + 0.1i$, (b) $\varepsilon_{y'} = -3 + 10i$ under $\theta_i = 60°$ (left) and $\theta_i = -60°$ (right).

**These results reveal that the thickness of the TAM absorber can be pushed to zero as the material loss (i.e. $\mathrm{Im}(\varepsilon_{y'})$) tends to be zero, which is totally different from the common understanding that thinner absorbers need larger material losses to maintain the same absorption. Actually, such extraordinary absorption originates from the enhanced electric fields normal to the air-TAM interface, which increase rapidly as $\mathrm{Im}(\varepsilon_{y'})$ decreases.**

It is worth noting that the physical origin of the extraordinary absorption by the ultrathin TAM is very similar to the extraordinary absorption by zero-index media [1, 2]. In addition, we notice that the absorption by the tilted (or asymmetric) hyperbolic media has been discussed in [3]. In these works, material loss is introduced to both $\varepsilon_{x'}$ and $\varepsilon_{y'}$, therefore, *the PIM is destroyed*. As a consequence, only when the material loss is small, can the near-perfect absorption be obtained. As long as the material loss is increased, the absorption will decrease and dramatic reflections will occur due to impedance mismatch.

## 7. Effective medium model of tilted CF array and ABE from dc to the GHz regime

In this section, we discuss the effective medium model of tilted conductive film (CF) array from dc to the GHz regime. The model we studied is illustrated by the left inset in Fig. S10. The CFs with a rotation angle of $\alpha$ are periodically aligned in an isotropic dielectric background (relative permittivity $\varepsilon_d$). The separation distance between two adjacent two CFs is $a$, which is much larger than the thickness $t$ of the CFs, i.e. $t \ll a$. In low frequency regime, the relative permittivity of the CFs can be expressed as $\varepsilon_{CF} = 1 + i \dfrac{Z_0/R_s}{k_0 t}$, where $R_s$ is sheet resistance of the CFs.

We assume that the separation distance $a$ is smaller than the wavelength $\lambda$ in the background dielectric, i.e. $a < \lambda$. Under this circumstance, we can approximately homogenize the tilted CF array based on classical Maxwell-Garnett theory [4] as an effective TAM (see the right inset in Fig. S10) with

$$\varepsilon_{x',\mathrm{eff}} = \frac{\varepsilon_{CF}\varepsilon_d a\cos\alpha}{\varepsilon_{CF}(a\cos\alpha - t) + \varepsilon_d t} \approx \varepsilon_d \quad \text{and} \quad \varepsilon_{y',\mathrm{eff}} = \frac{\varepsilon_d(a\cos\alpha - t) + \varepsilon_{CF} t}{a\cos\alpha} \approx \varepsilon_d + i\gamma(\omega), \tag{S17}$$

where $\gamma(\omega) = \dfrac{Z_0/R_s}{k_0 a\cos\alpha}$. Equation (S17) shows that $\varepsilon_{x',\mathrm{eff}} = \mathrm{Re}(\varepsilon_{y',\mathrm{eff}})$ is frequency-independent, while $\mathrm{Im}(\varepsilon_{y',\mathrm{eff}})$ relies on the working frequency.

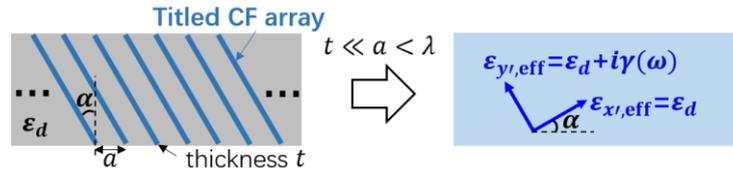

**Figure S10. Effective medium model of tilted CF array.** Illustrations of a tilted CF array in an isotropic dielectric background (left) and the corresponding effective medium model (right).

It is noteworthy that the Eq. (S17) is derived based on quasi-static limit, therefore, **there is no lower frequency limit**. Actually, with the titled CF array, we can realize PIM from the quasi-static limit to the GHz regime. As an example, we consider a tilted CF array with $a = 5$mm, $\varepsilon_d = 1$ and $\alpha = 30°$. Figure S11 presents the reflection coefficient at the interface of free space and the effective medium of the tilted CF array as functions of the incident angle and working frequency. The working frequency varies from dc to the KHz, MHz and GHz regimes. **The results clearly show that the PIM preserves from dc to the GHz regime, indicating the ultra-broadband ABE.**

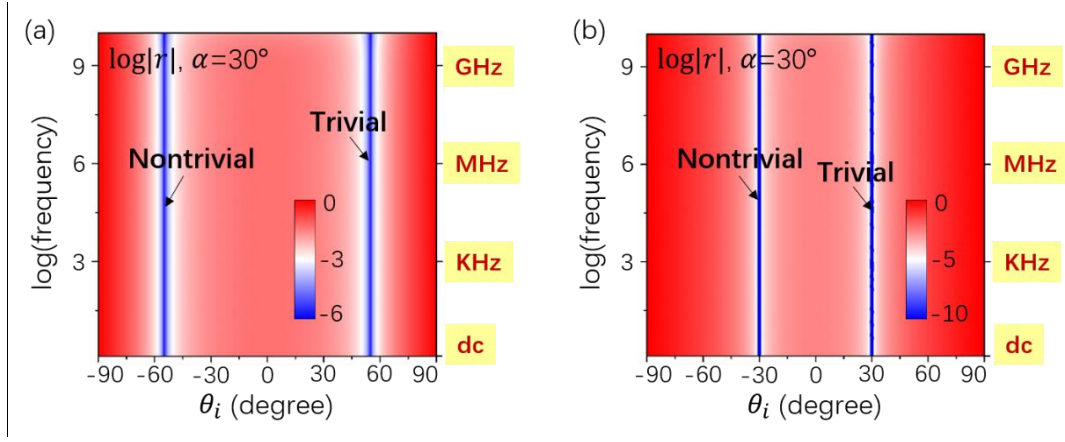

**Figure S11. ABE from dc to the GHz regime.** [(a) and (b)] Reflection coefficient at the interface of free space and the effective medium of the tilted CF array as functions of the incident angle and working frequency. The relevant parameters of the tilted CF array are (a) $a$=5mm, $\varepsilon_d = 2$ and $\alpha = 35.3°$, (b) $a$=5mm, $\varepsilon_d = 1$ and $\alpha = 30°$. The working frequency varies from dc to the GHz regime.

On the other hand, the upper frequency limit is determined by the separation distance $a$. The operating wavelength should satisfy $\lambda > a$ to guarantee the validity of the effective medium model. More specifically, we need to make sure that there are no diffraction waves, which requires $\frac{2\pi}{a} > \sqrt{\varepsilon_b}\frac{2\pi}{\lambda_0} + \sqrt{\varepsilon_b}\frac{2\pi}{\lambda_0}\sin\theta_i$ and $\frac{2\pi}{a} > \sqrt{\varepsilon_d}\frac{2\pi}{\lambda_0} + \sqrt{\varepsilon_b}\frac{2\pi}{\lambda_0}\sin\theta_i$, leading to

$$a < \frac{\lambda_0}{\sqrt{\varepsilon_b}(1+\sin\theta_i)} \quad \text{and} \quad a < \frac{\lambda_0}{\sqrt{\varepsilon_d}+\sqrt{\varepsilon_b}\sin\theta_i} \qquad (S18)$$

where $\lambda_0$ is the wavelength in free space.

**Since the separation distance of fabricated experimental samples is 5~10mm, the upper working frequency limit is around 20GHz, as implied by Eq. (S18). It is noteworthy that such upper frequency limit can be raised to higher frequency regimes (e.g. THz, infrared and optical regimes) by reducing the separation distance.**

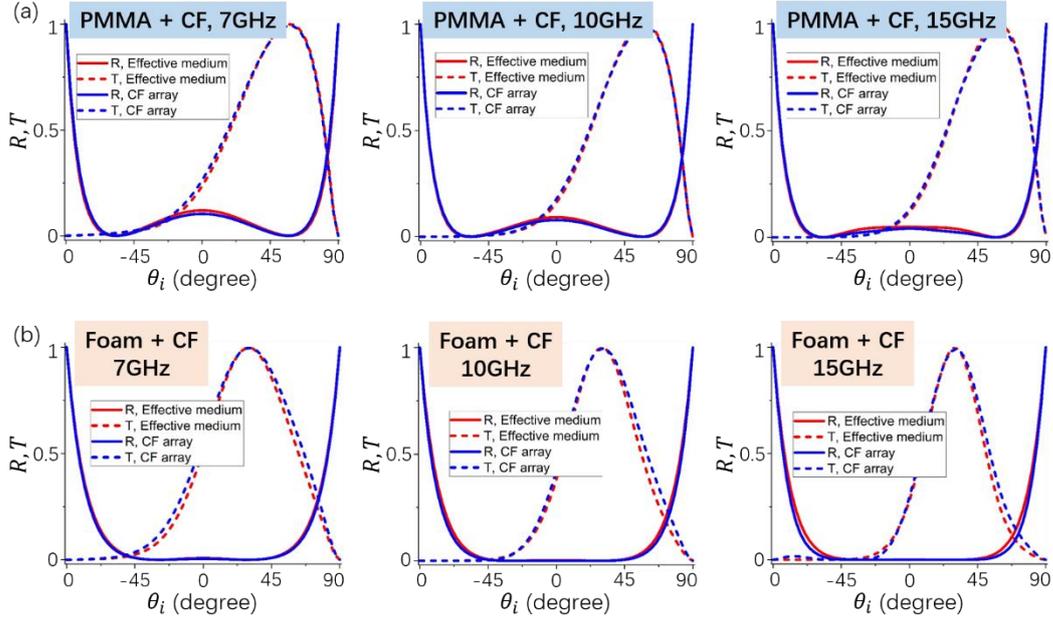

**Figure S12. Verification of the validity of effective medium models of experimental tilted CF array samples in the GHz regime.** [(a) and (b)] Reflectance and transmittance of the experimental sample (blue lines) and the corresponding effective medium (red lines) at 7GHz (left), 10GHz (middle) and 15GHz (right). In (a), the sample consisting of PMMA and ITO films is the same as that in Fig. 3. In (b), the sample consisting of foam and ITO films is same as that in Fig. 4.

Next, we check the validity of effective medium models of the experimental tilted CF array samples at studied frequencies 7GHz, 10GHz and 15GHz (i.e. the samples in Figs. 3 and 4 in the Main Text). First, we consider the sample consisting of PMMA and ITO films (the same sample in Fig. 3). The relevant parameters are $a=5$mm, $\varepsilon_b=1$, $\varepsilon_d=2.6$ and $\theta_i=58.2°$, indicating an upper frequency limit of 24.4GHz. Therefore, in our experiments, frequencies below 16GHz are studied. In Fig. S12(a), we compare the reflectance and transmittance for TM-polarized waves incident from air onto the experimental sample and the corresponding effective medium, showing quite good coincidence at 7GHz (left), 10GHz (middle) and 15GHz (right). Second, we consider the sample consisting of foam and ITO films (the same sample in Fig. 4), whose relevant parameters are $a=10$mm, $\varepsilon_b=\varepsilon_d=1$ and $\theta_i=30°$. This indicates an upper frequency limit of 20GHz. Figure S12(b) shows the reflectance and transmittance at 7GHz (left), 10GHz (middle) and 15GHz (right), also showing quite good coincidence between the experimental sample and the effective medium. **These results clearly demonstrate the validity of effective medium models of experimental samples at working frequencies (<16GHz).**

## 8. Optimal sheet resistance analysis, details of experimental samples and further experimental results

In this section, we first investigate the influences of sheet resistance $R_s$ in the attenuation rate of transmission waves in the TAM, and try to find out the optimal $R_s$ to obtain the largest attenuation rate, so that the absorbers can be as thin as possible. Then, we present more theoretical and experimental results of the tilted CF arrays with different $R_s$.

First, we study the tilted CF array with $\varepsilon_d \neq \varepsilon_b$ based on the effective TAM ( $\varepsilon_{x',\text{eff}} = \varepsilon_d$, $\varepsilon_{y',\text{eff}} = \varepsilon_d + i\gamma$ with $\gamma = (Z_0/R_s)/k_0 a \cos\alpha$ ). We know that under $\theta_i = -(\pi/2 - \alpha)$ with $\alpha = \arctan\left(\sqrt{\varepsilon_b}/\sqrt{\varepsilon_d}\right)$, we have $\gamma$-independent PIM and $\gamma$-controlled absorption of transmission waves in the TAM. **Generally, $\gamma$ cannot be too large or too small. Otherwise, the CFs tend to be PEC or air, thus there will be no absorption. Clearly, there exists an optimal value of $\gamma$ to obtain the largest absorption.** To study the optimal value, we derive the relation between $\text{Im}(k_y)$ (i.e. the imaginary part of $k_y$) and $\gamma$ under $\theta_i = -(\pi/2 - \alpha)$ as,

$$\left[\text{Im}(k_y)/k_0\right]^3 \left[\gamma^2 + (1+\varepsilon_d)^2\right]^2 + 4\left[\text{Im}(k_y)/k_0\right]^2 \gamma\sqrt{1+\varepsilon_d}\left[\gamma^2 + (1+\varepsilon_d)^2\right]$$
$$+ \left[\text{Im}(k_y)/k_0\right](1+\varepsilon_d)\left[\gamma^4 + \varepsilon_d^2(1+\varepsilon_d)^2 + \gamma^2(5 + 2\varepsilon_d + 2\varepsilon_d^2)\right] + 2\gamma(1+\varepsilon_d)^{3/2}(\gamma^2 + \varepsilon_d^2) = 0 \quad (S19)$$

According to Eq. (S19), the relation between $\text{Im}(k_y)$ and sheet resistance $R_s$ of the tilted CF array ( $a$ =5mm, $d$ =30mm) at 10GHz is plotted in Fig. S13(a). The red dots and black lines are related to CF arrays with different $\varepsilon_d$ and the effective media, respectively. It is seen that **as the increase of $\varepsilon_d$, both the optimal sheet resistance and the corresponding maximal value of $\left|\text{Im}(k_y)\right|$ decrease**. Moreover, we notice that **for the experimental sample consisting of PMMA ( $\varepsilon_d = 2.6$ ) and ITO films (i.e. the sample in Fig. 3 in the Main Text), the optimal sheet resistance is $\sim 0.35 Z_0$=132Ω for the CF array (or $\sim 0.31 Z_0$ =117Ω for the effective medium model)**. In experiments, it is not easy to fabricate the ITO films with the exact optimal sheet resistance. Actually, the ITO films we used possess a sheet resistance around 370Ω. In order to make the sheet resistance close to the optimal value, we have stacked two ITO films together, so that the superimposed ITO film has a half sheet resistance ~185Ω, as marked by the blue dashed line in Fig. S13(a). Figure S13(b) shows pictures of the fabricated sample and the ITO films.

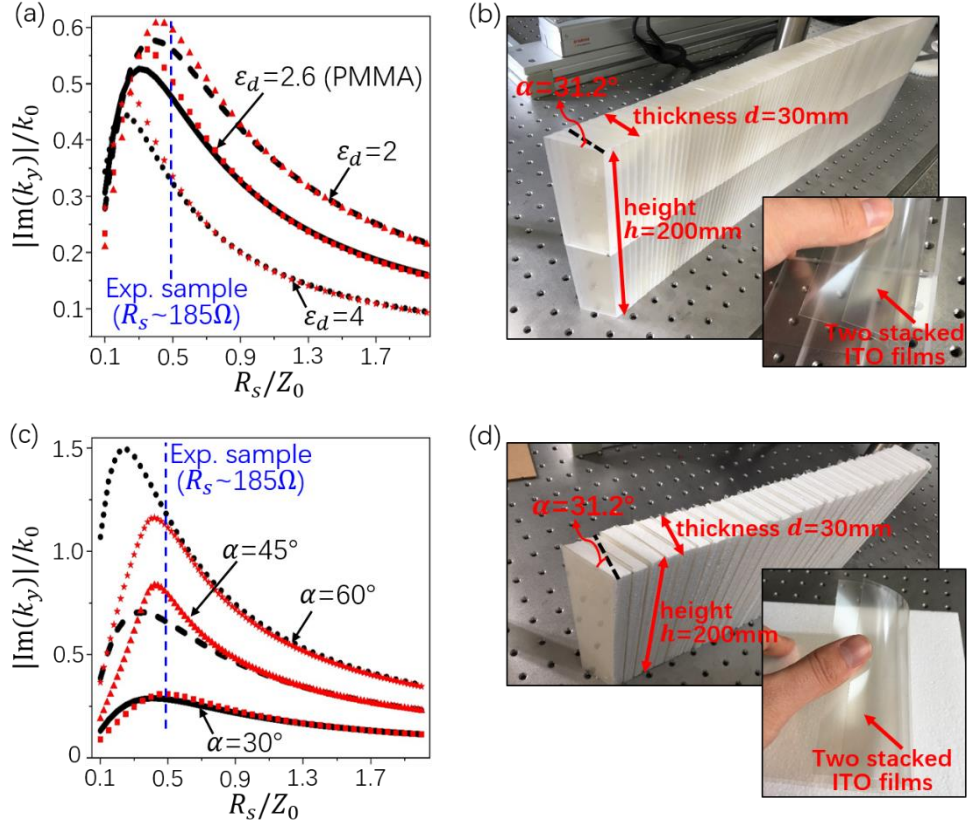

**Figure S13. Optimal sheet resistance analysis and details of experimental samples.** [(a) and (c)] $\text{Im}(k_y)$ as the function of sheet resistance $R_s$ of tilted CF arrays with (a) $\varepsilon_d \neq \varepsilon_b$, (c) $\varepsilon_d = \varepsilon_b$. In (a), the red dots and black lines are related to CF arrays with different $\varepsilon_d$ and the effective media, respectively. In (c), the red dots and black lines are related to CF arrays ($\varepsilon_d = 1$) with different $\alpha$ and the effective media, respectively. [(b) and (d)] Pictures of the fabricated samples and the used ITO films. Two ITO films are stacked together, so that the superimposed film has a sheet resistance around 185Ω. In (b), the sample is composed of PMMA and ITO films (the sample in Fig. 3). In (d), the sample is composed of foam and ITO films (the sample in Fig. 4).

Second, we study the tilted CF array with $\varepsilon_d = \varepsilon_b = 1$. Based on complex dispersion of the effective medium model, we obtain the $\text{Im}(k_y)$ under $\theta_i = -\alpha$ as,

$$\text{Im}(k_y)/k_0 = \frac{8\gamma \sin\alpha \sin 2\alpha}{8 + 3\gamma^2 + \gamma^2(4\cos 2\alpha + \cos 4\alpha)}. \tag{S20}$$

From Eq. (S20), we can find out the maximal $|\text{Im}(k_y)|$ as

$$\left[|\text{Im}(k_y)|/k_0\right]_{\max} = \sin\alpha \tan\alpha \tag{S21}$$

at the optimal $\gamma$ or sheet resistance

$$\gamma = 1 + \tan^2 \alpha \quad \text{or} \quad R_s = \frac{\cos\alpha}{k_0 a} Z_0. \tag{S22}$$

From Eq. (S22), we find out **the optimal sheet resistance of the experimental sample consisting of foam and ITO films (the sample in Fig. 4 in the Main Text) as $\sim 0.41 Z_0$ =155Ω**, as shown in Fig. S13(c). Similarly, we have stacked two ITO films together to construct a superimposed ITO film with $R_s \sim 185$Ω, as shown by pictures of the experimental sample and ITO films in Fig. S13(d). In addition, Eqs. (S20)-(S22) indicate an interesting thing, that is, **the maximal $\left|\text{Im}(k_y)\right|$ increases as the increase of $\alpha$, and tends to infinity as $\alpha$ goes to $90°$**, as shown in Fig. S13(c). This means that perfect absorption of electromagnetic/optical waves can be realized by using ultrathin tilted CF arrays with $\alpha \to 90°$ in an ultra-broad frequency band.

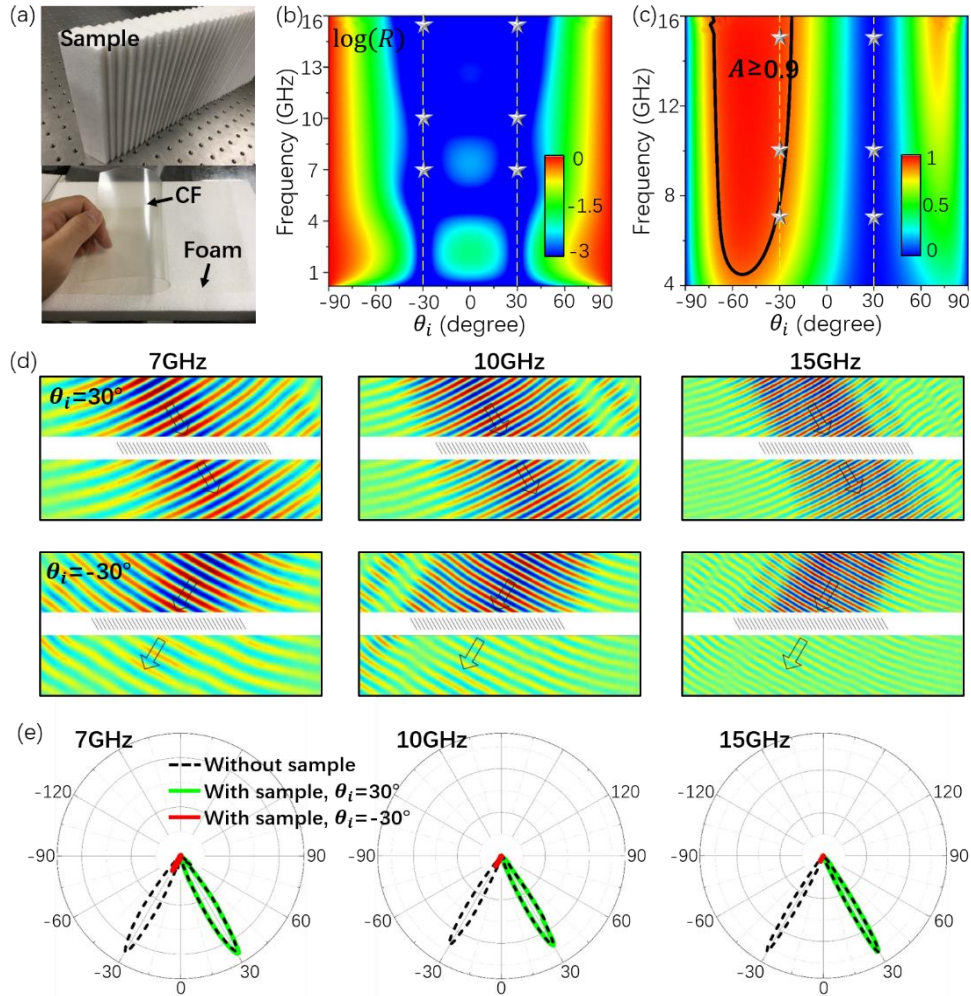

**Figure S14. Experimental demonstration of ultra-broadband PIM and near-perfect absorption by the tilted CF array (foam and ITO films) with $\alpha = 30°$, $a$ =10mm and $d$ =30mm. There is only one ITO film (~370Ω) between two adjacent foam layers.** (a) Photograph of the fabricated sample. Calculated (b) reflectance, (c) absorptance $A$ of the sample as functions of the $\theta_i$ and working

frequency. The stars denote the cases verified in experiments. (d) Measured near-field electric fields under $\theta_i = 30°$ (upper) and $\theta_i = -30°$ (lower) at 7GHz (left), 10GHz (middle) and 15GHz (right). (e) Measured far-field radiation power in the absence of sample (black dashed lines), with sample under $\theta_i = 30°$ (green lines) and $\theta_i = -30°$ (red lines) at 7GHz (left), 10GHz (middle) and 15GHz (right).

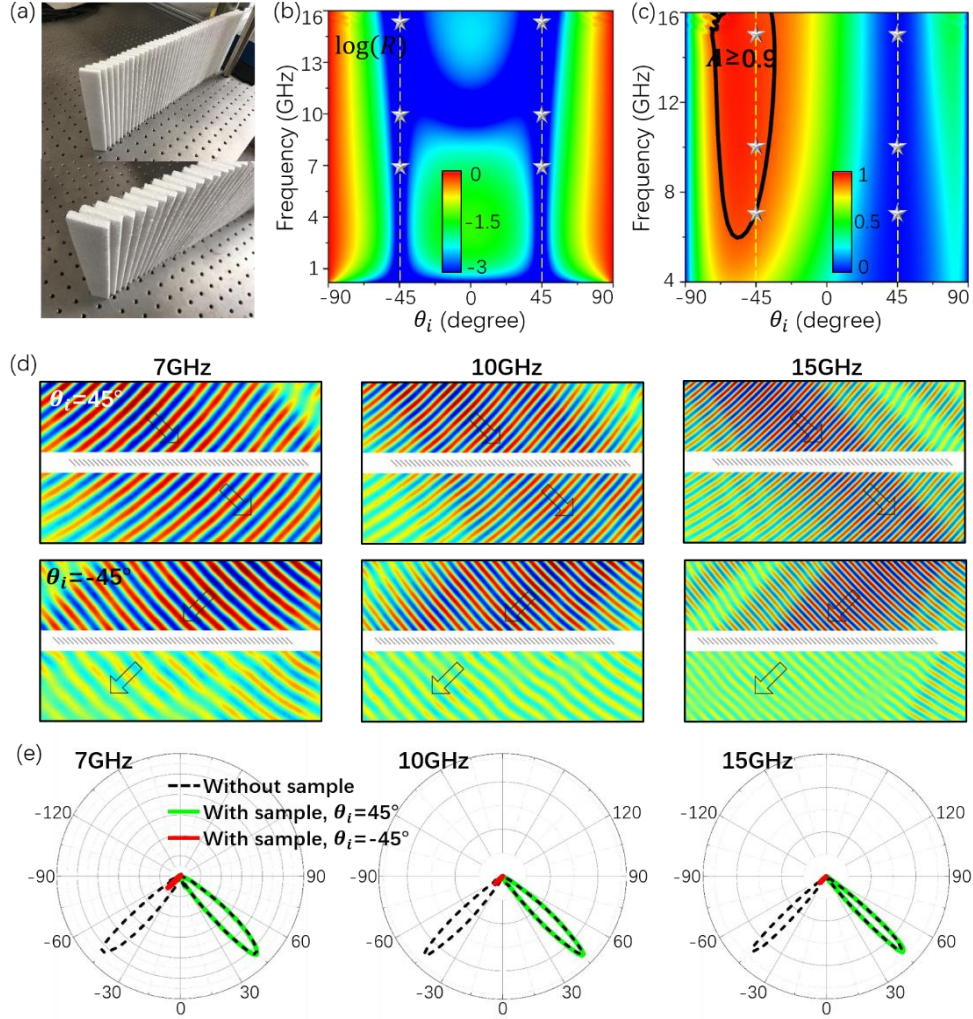

**Figure S15. Experimental demonstration of ultra-broadband PIM and near-perfect absorption by the tilted CF array (foam and ITO films) with $\alpha = 45°$, $a$ =10mm and $d$ =15mm. There is only one ITO film (~370Ω) between two adjacent foam layers.** (a) Photograph of the fabricated sample. Calculated (b) reflectance, (c) absorptance $A$ of the sample as functions of the $\theta_i$ and working frequency. The stars denote the cases verified in experiments. (d) Measured near-field electric fields under $\theta_i = 45°$ (upper) and $\theta_i = -45°$ (lower) at 7GHz (left), 10GHz (middle) and 15GHz (right). (e) Measured far-field radiation power in the absence of sample (black dashed lines), with sample under $\theta_i = 45°$ (green lines) and $\theta_i = -45°$ (red lines) at 7GHz (left), 10GHz (middle) and 15GHz (right).

Next, we have performed more numerical calculations and microwave experiments to show the influences of sheet resistance on the absorption performance. Figure S14 discusses the sample consisting of alternative foam and ITO films ($a$=10mm, $\alpha = 30°$), whose thickness and height are 30mm and 200mm, respectively. Different from the sample in Fig. 4 in the Main Text, here there is only one ITO film between two adjacent foam layers, as shown by pictures of the fabricated sample and ITO film in Fig. S14(a). Figures S14(b) and S14(c), respectively, present the simulated reflectance and absorptance on the fabricated sample, showing the zero reflection for all frequencies <16GHz under $\theta_i = \pm 30°$ and near-perfect absorption under $\theta_i = -30°$. These results have been confirmed by experimental measured electric fields under $\theta_i = 30°$ (upper) and $\theta_i = -30°$ (lower) at 7GHz (left), 10GHz (middle) and 15GHz (right) in Fig. S14(d). We have also measured the far-field radiation patterns (green and red lines) in Fig. S14(e). The black dashed lines denote the reference patterns in the absence of the sample. **These results show that the absorption is decreased a bit in comparison with the results in Fig. 4 in the Main Text, because the sheet resistance here is away from the optimal value (i.e. ~155Ω). Even so, ultra-broadband PIM and quite good absorption performance can still be observed, manifesting the robustness of wave absorption by the tilted CF arrays.**

Then, we change the rotation angle to $\alpha = 45°$, and reduce the thickness to 15mm. Figure S15(a) shows the fabricated sample. Figures S15(b) and S15(c), respectively, present the simulated reflectance and absorptance on the fabricated sample, showing the zero reflection for all frequencies <16GHz under $\theta_i = \pm 45°$ and near-perfect absorption under $\theta_i = -45°$. These results have been confirmed by experimental measured electric fields under $\theta_i = 45°$ (upper) and $\theta_i = -45°$ (lower) at 7GHz (left), 10GHz (middle) and 15GHz (right) in Fig. S15(d). We have also measured the far-field radiation patterns (green or red lines) in Fig. S15(e). The black dashed lines denote the reference patterns in the absence of the sample. **These results show that although the thickness is halved here, the absorption performance is even better than that in Fig. S15 with $\alpha = 30°$. Actually, as the rotation angle increases, ultra-broadband perfect absorption of electromagnetic waves can be realized even in deep-subwavelength CF arrays.**

## 9. Experimental setup and measurement methods

In experiments, we measure the near-field electric fields with the experimental setup shown in Fig. S16(a). An emitting horn antenna is placed ~0.8 meter away from the sample to generate the incident waves. A probing antenna is used to probe the near-field electric fields before and after the sample. The scanning rectangular area is of 700×150mm² (or 600×150mm²) before and after the sample consisting of PMMA (or foam) and ITO films. The scanning areas are on the central plane of the sample. The probing antenna is mounted to a computer controlled translational stage (not shown here). Both the probing antenna and the emitting horn antenna are connected to a network analyzer (KEYSIGHT PNA Network Analyzer N5224B) for data acquisition.

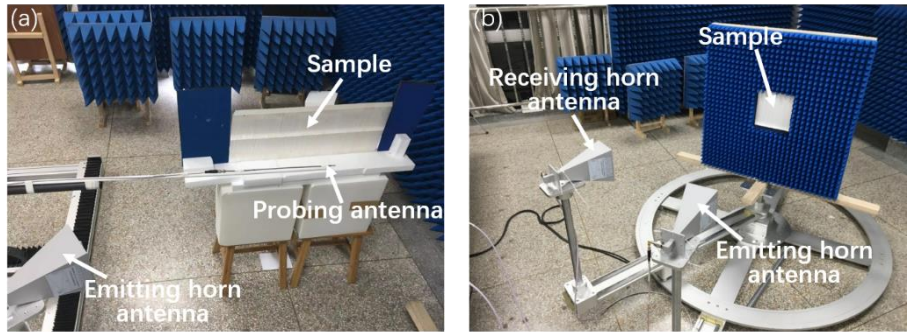

**Figure S16. Experimental setup and measurement methods.** Pictures of experimental setup for the measurement of (a) near-field electric fields, (b) far-field power radiation patterns.

We note that only the electric fields perpendicular to the propagation direction are measured because of the selectivity of the probing antenna in measurement. Therefore, in the scan area before the sample, the measured electric fields come from the incident waves and a part of the possible reflection waves except of the special case of $\theta_i = \pm 45°$. **In order to further confirm the zero reflection and near-perfect absorption in experiments, we have also measured the electric fields in the absence of samples and the far-field power radiation patterns (see experimental setup in Fig. S16(b))**. Regarding to the experiments in Fig. 3 in the Main Text, we have also measured near-field electric fields in the absence of samples under $\theta_i = 58.2°$ at 7GHz (left), 10GHz (middle) and 15GHz (right), as shown in Fig. S17. Compared with the field distributions with samples in Fig. 3, we can that the electric fields before the sample are almost the same. This demonstrates that there are almost no reflection waves in the existence of samples (Fig. 3), thus confirming the ultra-broadband PIM and near-perfect absorption. Regarding to the experiments in Fig. 4 in the Main Text, we have measured far-field power radiation patterns under $\theta_i = 30°$ (green lines) and $\theta_i = -30°$ (red lines) at 7GHz (left), 10GHz (middle) and 15GHz (right), as shown in Fig. S18. The black dashed lines denote the reference patterns in the absence of the sample. These results clearly demonstrate the ultra-broadband PIM and near-perfect absorption.

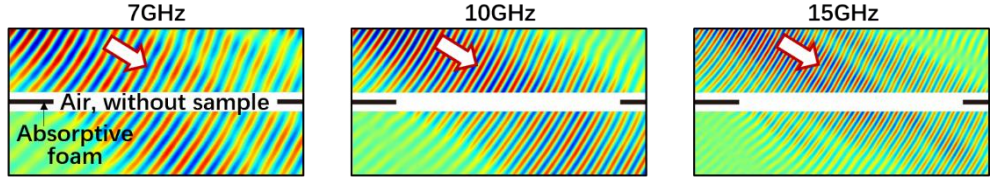

**Figure S17. Measured electric fields in the absence of samples for a reference (regarding to the sample in Fig. 3).** Measured near-field electric fields in the absence of samples under the incident angle of $\theta_i = 58.2°$ at 7GHz (left), 10GHz (middle) and 15GHz (right)

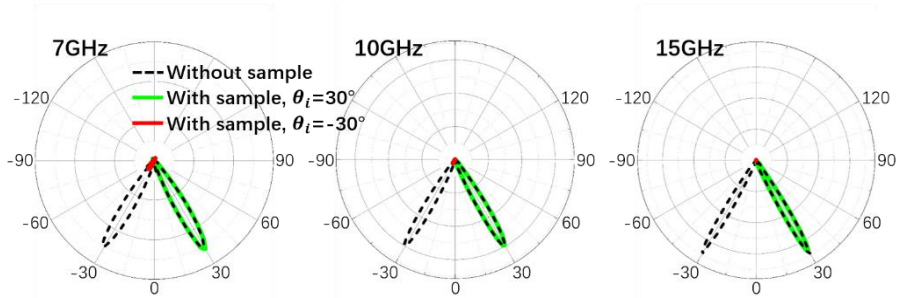

**Figure S18. Measured far-field power radiation patterns (regarding to the sample in Fig. 4).** Measured far-field power radiation patterns for the sample consisting of foam and ITO films under $\theta_i = 30°$ (green lines) and $\theta_i = -30°$ (red lines) at 7GHz (left), 10GHz (middle) and 15GHz (right). The black dashed lines denote the reference patterns in the absence of samples.

In the far-field experiments, the power radiation pattern of the sample is measured with the experimental setup shown in Fig. S17(b). An emitting horn antenna is placed 1m away from the sample to generate the incident waves. A receiving horn antenna placed at the same distance is used to measure the radiation pattern. The receiving horn antenna can be freely moved around the sample so that we could receive scattering signals in all directions. Both the emitting and receiving horn antennas are connected to a vector network analyzer (KEYSIGHT PNA Network Analyzer N5224B) for data acquisition. The power radiation pattern in the absence of sample is measured as a reference.

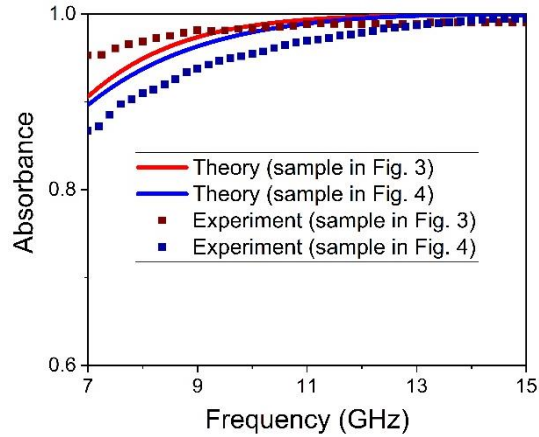

**Figure R19. Absorptance of the experimental samples within the measuring frequency range 7-15GHz.** The red and blue lines/dots are related to the sample in Fig. 3 (incident angle -30°) and the sample in Fig. 4 (incident angle -58.2°), respectively. The lines and dots denote the theoretical and experimental results, respectively.

Furthermore, through integrating far-field power for all directions, the absorbance by the designed absorbers can be evaluated. Figure S19 presents the absorptance distributions of the experimental sample in Fig. 3 (incident angle -30°) and the experimental sample in Fig. 4 (incident angle -58.2°), as shown by the red and blue lines, respectively. The experimental results (dots) are in good coincidence with the theoretical results (lines). The absorption is quite high within the measuring frequency range 7-15GHz. The relatively low absorption at low frequencies attributes to the long wavelengths. Through increasing the sample thickness, the absorption performance can be greatly improved.

## 10. Improvement of absorption by using reflectors

In the Main Text, the absorption by the tilted CF array is asymmetric with respect to $\theta_i$, and is small under $\theta_i > 0$. Here, we show that the absorption can be greatly improved by using a reflector. Figure S20 (a) presents the absorptance by the sample in Fig. 3 in the Main Text, but with a PEC reflector behind. Interestingly, symmetric high absorption (>0.9) in an ultra-broad frequency band (4~16GHz) under $\theta_i = \pm 58.2°$ is seen, which is further confirmed by the simulations in Fig. S20(b). The similar symmetric high absorption is also observed by the sample in Fig. 4 in the Main Text with a PEC reflector behind, as shown in Figs. S20(c) and S20(d). **The results clearly show that with a PEC reflector, symmetric absorption can be obtained, and both the angular performance and band width of absorption can be greatly improved.**

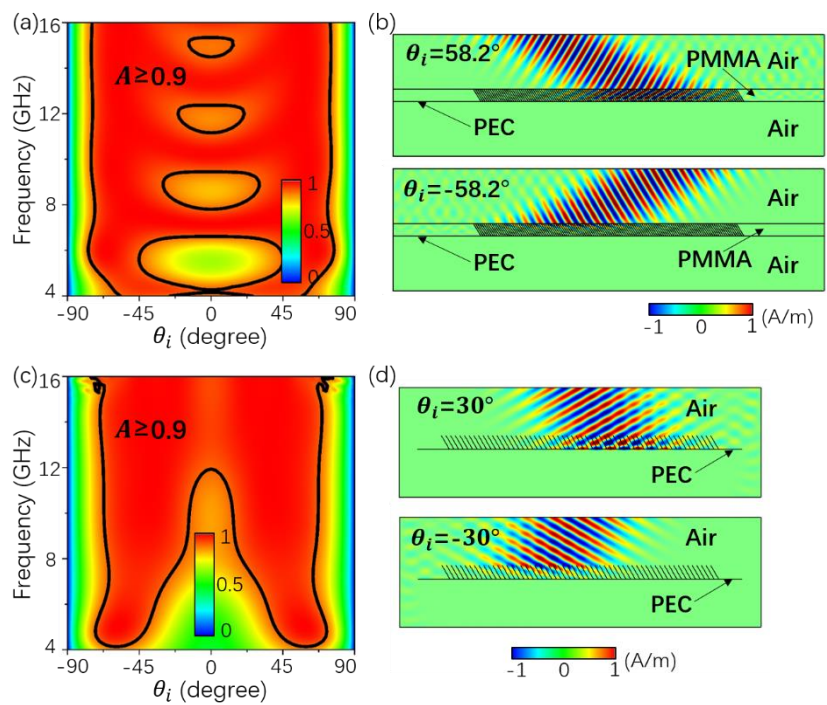

**Figure S20. Improving absorption by using PEC reflectors.** (a) Absorptance by the experimental sample (PMMA and ITO films) with a PEC reflector as functions of $\theta_i$ and working frequency. (b) Simulated magnetic field-distributions under $\theta_i = \pm 58.2°$ at 10GHz. The sample in (a) and (b) is the same as that in Fig. 3. (c) Absorptance by the experimental sample (foam and ITO films) with a PEC reflector as functions of $\theta_i$ and working frequency. (b) Simulated magnetic field-distributions under $\theta_i = \pm 30°$ at 10GHz. The sample in (c) and (d) is the same as that in Fig. 4.